\documentclass[a4paper,11pt]{article}
\pdfoutput=1 
\usepackage{aas_macros,braket}
\usepackage{jcappub,natbib} 
\usepackage[T1]{fontenc} 

\title{Minimum variance estimation of statistical anisotropy via galaxy survey}

\author[a]{Maresuke Shiraishi,}
\author[b,c]{Teppei Okumura,}
\author[c]{and Kazuyuki Akitsu}

\affiliation[a]{Department of General Education, National Institute of Technology, Kagawa College, 355 Chokushi-cho, Takamatsu, Kagawa 761-8058, Japan}
\affiliation[b]{Academia Sinica Institute of Astronomy and Astrophysics (ASIAA), No. 1, Section 4, Roosevelt Road, Taipei 10617, Taiwan}
\affiliation[c]{Kavli Institute for the Physics and Mathematics of the Universe (WPI), UTIAS, The University of Tokyo, Chiba 277-8583, Japan}

\emailAdd{mare@nagoya-u.jp}
\emailAdd{tokumura@asiaa.sinica.edu.tw}
\emailAdd{kazuyuki.akitsu@ipmu.jp}

\abstract{
We consider the benefits of measuring cosmic statistical anisotropy from redshift-space correlators of the galaxy number density fluctuation and the peculiar velocity field without adopting the plane-parallel (PP) approximation. Since the correlators are decomposed using the general tripolar spherical harmonic (TripoSH) basis, we can deal with wide-angle contributions untreatable by the PP approximation, and at the same time, target anisotropic signatures can be cleanly extracted. We, for the first time, compute the covariance of the TripoSH decomposition coefficient and the Fisher matrix to forecast the detectability of statistical anisotropy. The resultant expression of the covariance is free from nontrivial mixings between each multipole moment caused by the PP approximation and hence the detectability is fully optimized. Compared with the analysis under the PP approximation, the superiority in detectability is always confirmed, and it is highlighted, especially in the cases that the shot noise level is large and that target statistical anisotropy has a blue-tilted shape in Fourier space. The application of the TripoSH-based analysis to forthcoming all-sky survey data could result in constraints on anisotropy comparable to or tighter than the current cosmic microwave background ones.
}

\begin{document}



\maketitle
\flushbottom

\section{Introduction}

The assumption of global isotropy of the Universe is one key underpinning of the concordance $\Lambda$CDM cosmology. This is linked with the cosmic no-hair conjecture at the primordial inflationary stage. On the other hand, recently, possibilities of the departure from isotropy have been widely and thoroughly argued. The presence of some source fields as vector or higher-spin fields (e.g.~\cite{Ratra:1991bn,Soda:2012zm,Maleknejad:2012fw,Kehagias:2017cym}), an inflating solid (e.g.~\cite{Bartolo:2013msa, Bartolo:2014xfa}), fossil gravitational waves (e.g.~\cite{Jeong:2012df,Dai:2013kra}) and large-scale tides (e.g.~\cite{Akitsu:2016leq,Akitsu:2017syq,Li:2017qgh,Chiang:2018mau,Akitsu:2019avy}) in both early and late Universe has been proposed and studied for such possibilities.

Broken isotropy or equivalently broken rotational invariance imprints quite unique signatures in the two-point correlation function (2PCF) or the power spectrum between a variety of cosmic observables such as the cosmic microwave background (CMB), large-scale structure, the 21-cm radiation and the gravitational wave background. The anisotropic signatures have been tested with diverse observed data, however, there is no detection up to now. In the analysis using angular power spectra computed on the 2D sphere, a nonvanishing signal in off-diagonal multipole modes is an observational indicator. No detection of it places upper bound on the anisotropic component. The stringent one comes from the CMB, reading $\sim 10^{-2}$ of the isotropic component \cite{Ade:2015lrj,Ade:2015hxq,Akrami:2018odb,Shaikh:2019dvb}. Such constraints can be utilized for determining the particle content of the Universe (e.g.~\cite{Bartolo:2015dga,Bartolo:2017sbu,Akrami:2019izv}).

On the other hand, very recently, the isotropy test in the 3D space has also been thoroughly examined. In ref.~\cite{Shiraishi:2016wec}, we developed an efficient way to extract the anisotropic signal from the 2PCF of 3D galaxy clustering. The current upper bound obtained in this way is within an order of magnitude of the CMB constraint \cite{Sugiyama:2017ggb}, while a drastic update is expected in the future since the signal-to-noise ratio in the 3D analysis grows even faster than the 2D one as the number of accessible Fourier modes increases \cite{Shiraishi:2016wec,Bartolo:2017sbu,Akitsu:2019avy}.

The observational constraints and their forecasts mentioned above were obtained imposing the plane-parallel (PP) approximation where two different line-of-sight (LOS) directions $\hat{s}_1$ and $\hat{s}_2$ are identified with each other. This approximation has worked well in the analysis of galaxy surveys so far because the data where the visible angle of the survey area become too large has not been used. In contrast, this approximation is not applicable to the analysis including any wide-angle contribution targeted by proposed all-sky surveys such as SPHEREx \cite{Dore:2014cca}, Euclid \cite{Laureijs:2011gra} and WFIRST \citep{Spergel:2013tha}. Establishing a complete analysis methodology without the PP approximation is therefore a pressing issue.%
\footnote{See refs.~\cite{Matsubara:1999cf,Pope:2004cc,Okumura:2007br} for a few studies on analyzing observed galaxy clustering in configuration space.}

In the absence of the PP approximation, the 2PCF is characterized by three directions: two LOS vectors, $\hat{s}_1$ and $\hat{s}_2$, and its separation vector, $\hat{s}_{12} \equiv \widehat{{\bf s}_1 - {\bf s}_2}$, where hat denotes a unit vector. In the appendix of ref.~\cite{Shiraishi:2016wec}, we found that the directional dependence can be completely decomposed using the general tripolar spherical harmonic (TripoSH) basis $\{ Y_{\ell}(\hat{s}_{12}) \otimes \{ Y_{\ell_1}(\hat{s}_1) \otimes Y_{\ell_2}(\hat{s}_2) \}_{\ell'} \}_{LM}$ \cite{Varshalovich:1988ye}. If the Universe is isotropic, nonvanishing coefficients are confined to $L = 0$. There are many previous works on the response of the $L = 0$ coefficients on a variety of isotropic Universe models (e.g.~\cite{Szalay:1997cc,Szapudi:2004gh,Papai:2008bd,Bertacca:2012tp,Yoo:2013zga,Raccanelli:2013dza}).%
\footnote{See e.g., refs.~\cite{Reimberg:2015jma,Castorina:2017inr,Beutler:2018vpe,Castorina:2019hyr,Taruya:2019xsf} for other decomposition approaches.}
In contrast, ref.~\cite{Shiraishi:2016wec} showed that the $L > 0$ coefficients become a clean indicator of broken isotropy.

As for $L = 0$, very recently, the covariance of the TripoSH coefficient was also examined in our paper~\cite{Shiraishi:2020nnw}. We then found that nontrivial mixings between different multipole moments as seen in the PP-limit covariance do not exist. This fact minimizes the covariance and hence drastically optimizes the signal-to-noise ratio, especially at higher multipole moments.

In this paper, in anticipation of the anisotropic signal measurement using wide-angle data, we generalize our covariance formalism in ref.~\cite{Shiraishi:2020nnw} by including $L > 0$. As in ref.~\cite{Shiraishi:2020nnw}, we consider power spectra of density and velocity fields in redshift space as observables in galaxy redshift surveys, peculiar velocity surveys and kinematic Sunyaev-Zel'dovich surveys (e.g.~\cite{Sunyaev:1980nv,Kaiser:1987qv,Strauss:1995fz,Burkey:2003rk,Hand:2012ui,Okumura:2013zva,Koda:2013eya,Sugiyama:2015dsa}). We then confirm that, even for $L >0$, the covariance is minimized by virtue of high separability of the general TripoSH basis. We also compute the Fisher matrix and forecast the constraints on a widely-used isotropy breaking parameter $g_{LM}$ [see eqs.~\eqref{eq:Pm_GLM} and \eqref{eq:gLM_def} for definition] from the galaxy density and velocity fields. It is then found that our new TripoSH-based analysis always surpasses the previous PP-limit one in detectability of $g_{LM}$ thanks to minimizing the covariance, and the superiority becomes remarkable if the shot noise level enlarges or if target anisotropy has a blue-tilted shape in Fourier space.

This paper is organized as follows. In the next section, we summarize the linear-order expressions of the galaxy number density fluctuation and the peculiar velocity field underlying our discussions. In section~\ref{sec:TripoSH}, we perform the TripoSH decomposition of the 2PCF of the density and velocity fields and estimate the signal arising from the anisotropic Universe models and the covariance of the TripoSH coefficient. The error estimation on $g_{LM}$ is done in section~\ref{sec:error}. The final section is devoted to the conclusion of this paper. In appendix~\ref{appen:LPP}, the Fisher matrix computation in the PP limit based on the bipolar spherical harmonic (BipoSH) decomposition is argued. In appendix~\ref{appen:math}, some mathematical identities utilized in this paper are summarized.

\section{Linear theory for cosmic density and velocity fields} \label{sec:linear}

In this paper, as cosmic observables, we take into account two scalar quantities: the galaxy number density contrast $\delta({\bf s})  \equiv n({\bf s}) / \bar{n}(s) - 1 $ and the LOS peculiar velocity field $u({\bf s}) = {\bf v}({\bf s}) \cdot \hat{s}$ in redshift space. Since our main interest is to extract large-scale information on statistical anisotropy from galaxy clustering, we may work with the linear theory representation \cite{Hamilton:1997zq,Burkey:2003rk,Yoo:2013zga}:
\begin{equation}
  X({\bf s}) = \int \frac{d^3 k}{(2\pi)^3} e^{i {\bf k} \cdot {\bf s}}
  F^X({\bf k}, \hat{s}) , \label{eq:X_s}
\end{equation}
where $X = \{ \delta, u\}$ and 
\begin{eqnarray}
  \begin{split}
  F^\delta({\bf k}, \hat{s}) &\equiv \left[b  - i \frac{\alpha}{ks} (\hat{k} \cdot \hat{s}) f + (\hat{k} \cdot \hat{s})^2 f  \right] \delta_m({\bf k}) 
, \\
  F^u({\bf k}, \hat{s})  &\equiv i \frac{aH}{k} (\hat{k} \cdot \hat{s}) f \delta_m ({\bf k}) .
  \end{split}
\end{eqnarray}
The linear bias parameter $b$, the linear growth rate $f$, the scale factor $a$, the Hubble parameter $H$, the selection function $ \alpha \equiv d \ln \bar{n}(s) / d \ln s +  2$, the real-space matter density fluctuation $\delta_m({\bf k})$ and $F^X({\bf k}, \hat{s})$ depend on time, redshift or the conformal distance although it is not clearly stated as an argument for notational convenience. This convention is also adopted henceforth unless the parameter dependence is nontrivial. Note that $F^X({\bf k}, \hat{s})$ does not correspond to the Fourier counterpart of $X({\bf s})$ because there still remains the $\hat{s}$ dependence. For later convenience, we expand the angular dependence due to the redshift-space distortion using the Legendre polynomials ${\cal L}_\ell(x)$ as
\begin{equation}
  F^X({\bf k}, \hat{s}) = \sum_{j} c_j^X(k) {\cal L}_j(\hat{k} \cdot \hat{s}) \delta_m({\bf k}) ,
  \end{equation}
where $c_j^{X  *} = (-1)^j c_j^X$ and
\begin{eqnarray}
  \begin{split}
    & c_0^\delta = b + \frac{1}{3} f   , \ \ 
  c_1^\delta = - i \frac{\alpha}{k s} f , \ \ 
  c_2^\delta = \frac{2}{3} f , \ \ 
  c_{j \geq 3}^{\delta} = 0 , \\
  & c_1^u = i \frac{aH}{k} f, \ \ 
  c_0^u = c_{j \geq 2}^u = 0 .
  \end{split} \label{eq:delta_u_coeff}
\end{eqnarray}

Now, we assume that the matter distribution in real space is statistically homogeneous but is allowed to be statistically anisotropic in anticipation of anisotropic cosmological scenarios. The matter power spectrum then takes the form
\begin{equation}
 \Braket{ \delta_m({\bf k}_1) \delta_m({\bf k}_2) } = (2\pi)^3 \delta^{(3)}({\bf k}_1 + {\bf k}_2) P_m({\bf k}_1).
\end{equation}
In this case, the 2PCF of the density and velocity fields is written as 
\begin{eqnarray}
 \xi^{X_1 X_2}({\bf s}_{12}, \hat{s}_1, \hat{s}_2) &\equiv& \Braket{X_1({\bf s}_1) X_2({\bf s}_2)} \nonumber \\ 
 &=& \int \frac{d^3 k}{(2\pi)^3} e^{i {\bf k} \cdot {\bf s}_{12}} P^{X_1 X_2}({\bf k}, \hat{s}_1, \hat{s}_2) , \label{eq:xi_homo}
\end{eqnarray}
where ${\bf s}_{12} \equiv {\bf s}_1 - {\bf s}_2$ and 
\begin{equation} 
  P^{X_1 X_2}({\bf k}, \hat{s}_1, \hat{s}_2)
  = \sum_{j_1 j_2} c_{j_1}^{X_1}(k) (-1)^{j_2}c_{j_2}^{X_2} (k)
  {\cal L}_{j_1}(\hat{k} \cdot \hat{s}_1) 
  {\cal L}_{j_2}(\hat{k} \cdot \hat{s}_2) P_m({\bf k}) . \label{eq:P_homo}
\end{equation}
Note that $P^{X_1 X_2}({\bf k}, \hat{s}_1, \hat{s}_2)$ cannot be regarded as the Fourier counterpart of $\xi^{X_1 X_2}({\bf s}_{12}, \hat{s}_1, \hat{s}_2)$ because the $\hat{s}_1$ and $\hat{s}_2$ dependence still leaves.

Our formalism developed below is based on these formulae. Although the contributions of unequal-time correlators are not included for simplicity, they become treatable through a small extension. The above variables depend on cosmological parameters. Our numerical analysis performed below is done by fixing their values to be consistent with the latest CMB limits \cite{Aghanim:2018eyx}.

\section{Tripolar spherical harmonic decomposition} \label{sec:TripoSH}

In this section, we decompose the 2PCF \eqref{eq:xi_homo} using the TripoSH basis, and extract the anisotropic signatures. Moreover, we compute the covariance of the decomposition coefficient.

\subsection{Decomposition rule}

As seen in eq.~\eqref{eq:xi_homo}, the 2PCF is characterized by $\hat{s}_{12}$, $\hat{s}_1$ and $\hat{s}_2$. For a general angular decomposition basis, let us introduce the TripoSH function taking these three angles as arguments \cite{Varshalovich:1988ye,Shiraishi:2016wec}:
\begin{eqnarray}
  {\cal X}_{\ell \ell_1\ell_2 \ell'}^{LM}(\hat{s}_{12},\hat{s}_1,\hat{s}_2)
  &\equiv& \{ Y_{\ell}(\hat{s}_{12}) \otimes \{ Y_{\ell_1}(\hat{s}_1) \otimes Y_{\ell_2}(\hat{s}_2) \}_{\ell'} \}_{LM} \nonumber \\
  &=& \sum_{m m_1 m_2 m' } {\cal C}_{\ell m \ell' m'}^{LM} {\cal C}_{\ell_1 m_1 \ell_2 m_2}^{\ell' m'} 
  Y_{\ell m}(\hat{s}_{12}) Y_{\ell_1 m_1}(\hat{s}_1) Y_{\ell_2 m_2}(\hat{s}_2) , \label{eq:Xbasis_def}
\end{eqnarray}
where ${\cal C}_{l_1 m_1 l_2 m_2}^{l_3 m_3} \equiv (-1)^{l_1 - l_2 + m_3} \sqrt{2l_3 + 1} \left( \begin{matrix} l_1 & l_2 & l_3 \\ m_1 & m_2 & -m_3 \end{matrix}  \right)$ is the Clebsch-Gordan coefficient. We decompose the 2PCF according to
\begin{equation}
  \xi^{X_1 X_2}({\bf s}_{12}, \hat {s}_1, \hat{s}_2) 
  = \sum_{\ell\ell_1\ell_2 \ell' LM} \Xi_{\ell\ell_1\ell_2 \ell'}^{LM X_1 X_2}(s_{12}) 
     {\cal X}_{\ell\ell_1\ell_2\ell'}^{LM}(\hat{s}_{12},\hat{s}_1,\hat{s}_2) . \label{eq:TripoSH_Xi_def}
\end{equation}
Note that the coefficient with $L = M = 0$, $\Xi_{\ell\ell_1\ell_2\ell}^{00 X_1 X_2}$, is equivalent to $\Xi_{\ell\ell_1\ell_2}^{X_1 X_2}$ in ref.~\cite{Shiraishi:2020nnw}. As confirmed in ref.~\cite{Shiraishi:2016wec} and the next subsection of this paper, distinctive signatures of statistical anisotropy appear for $L > 0$.

Now, we consider the TripoSH decomposition of eq.~\eqref{eq:xi_homo}. For later convenience, let us expand $P^{X_1 X_2}$ according to
  \begin{equation}
P^{X_1 X_2}({\bf k}, \hat{s}_1, \hat{s}_2) 
 = \sum_{\ell \ell_1 \ell_2 \ell' LM} \Pi_{\ell \ell_1 \ell_2 \ell'}^{LM X_1 X_2}(k) 
  {\cal X}_{\ell \ell_1\ell_2 \ell'}^{LM}(\hat{k},\hat{s}_1,\hat{s}_2). \label{eq:TripoSH_P_def}
  \end{equation}
  Plugging this into eq.~\eqref{eq:xi_homo} and simplifying the $\hat{k}$ integral by use of eqs.~\eqref{eq:math_expand} and \eqref{eq:math_Ylm}, we find that $\xi^{X_1 X_2}$ recovers eq.~\eqref{eq:TripoSH_Xi_def} with 
\begin{equation}
  \Xi_{\ell\ell_1\ell_2\ell'}^{LM X_1 X_2}(s_{12}) 
  = i^{\ell} \int_0^\infty \frac{k^2 dk}{2\pi^2}    j_{\ell}(k s_{12})
  \Pi_{\ell \ell_1 \ell_2 \ell'}^{LM X_1 X_2}(k) . \label{eq:hankel}
\end{equation}
The practical form of $\Pi_{\ell \ell_1 \ell_2 \ell'}^{LM X_1 X_2}$ is obtained computing
\begin{equation}
  \Pi_{\ell \ell_1 \ell_2 \ell'}^{LM X_1 X_2}(k) 
  =  \int d^2 \hat{k}  \int d^2 \hat{s}_1  \int d^2 \hat{s}_2 \,
  P^{X_1 X_2}({\bf k}, \hat{s}_1, \hat{s}_2) {\cal X}_{\ell \ell_1\ell_2 \ell'}^{LM *}(\hat{k},\hat{s}_1,\hat{s}_2) . \label{eq:Pi_def}
\end{equation}

It is convenient to introduce a reduced coefficient:
\begin{equation}
  {\cal P}_{\ell \ell_1 \ell_2 \ell'}^{LM X_1 X_2}(k) 
  \equiv (-1)^{L + \ell'} \frac{h_{\ell_1 \ell_2 \ell'} h_{\ell \ell' L} }{\sqrt{4\pi (2\ell' + 1)}}
  \Pi_{\ell \ell_1 \ell_2 \ell'}^{LM X_1 X_2}(k), \label{eq:calP_def}
\end{equation}
where
\begin{equation}
h_{l_1 l_2 l_3} \equiv \sqrt{\frac{(2 l_1 + 1)(2 l_2 + 1)(2 l_3 + 1)}{4 \pi}} \left(\begin{matrix}
  l_1 & l_2 & l_3 \\
   0 & 0 & 0 
\end{matrix}\right).
\end{equation}
The coefficient ${\cal P}_{\ell \ell_1 \ell_2 \ell'}^{LM X_1 X_2}$ corresponds to the BipoSH coefficient in the PP limit $\hat{s}_1 = \hat{s}_2$ (see appendix~\ref{appen:LPP_BipoSH} for definition) with a simple relation: 
\begin{equation}
  P_{\ell \ell'}^{LM X_1 X_2}(k) = \sum_{\ell_1 \ell_2} {\cal P}_{\ell \ell_1 \ell_2 \ell'}^{LM X_1 X_2}(k) ; \label{eq:calP_2_P}
\end{equation}
thus, facilitates the comparison with the previous PP-limit results. One can see from this that, in the PP approximation, the two multipole moments $\ell_1$ and $\ell_2$, which are associated with $\hat{s}_1$ and $\hat{s}_2$, respectively, are contracted and disappear. This gives rise to the information loss (see appendix~\ref{appen:LPP_covmat} for details).

\subsection{Anisotropic signal}

If the matter distribution in real space breaks statistical isotropy, there remains $\hat{k}$ dependence in its power spectrum. We express it with a generic form,
\begin{equation}
  P_m({\bf k}) = \bar{P}_m(k) \sum_{LM} G_{L M}(k) Y_{LM}(\hat{k}) , \label{eq:Pm_GLM}
\end{equation}
where $\bar{P}_m \in \mathbb{R}$ quantifies the isotropic component (and hence $G_{00} = \sqrt{4\pi}$), and $G_{L \geq 1, M}(k)$ represents the fraction of the departure from isotropy, obeying $G_{LM} = (-1)^{M} G_{L, -M}^* \in \mathbb{C}$ and $G_{L={\rm odd}, M} = 0$ from $P_m({\bf k}) = P_m(-{\bf k}) = P_m^*(-{\bf k})$. This type of directional dependence can originate from the primordial curvature power spectrum, e.g., if higher spin fields couple to some scalar fields in the inflationary era. In the simplest case where there are spin-1 vectors coupled to inflaton scalars, $G_{2M}$ does not vanish as well as $G_{00}$ (e.g.~\cite{Watanabe:2010fh,Bartolo:2012sd,Ohashi:2013qba,Bartolo:2014hwa,Naruko:2014bxa,Bartolo:2015dga,Abolhasani:2015cve}). More generally, spin-$s$ fields generate $G_{00}$, $G_{2M}$, $G_{4M}$, $\cdots$, $G_{2(s-1), M}$ and $G_{2s, M}$ \cite{Kehagias:2017cym,Bartolo:2017sbu}. Nonvanishing $G_{L>2, M}$ can also be sourced from two-form fields \cite{Obata:2018ilf} and even from spin-1 vectors coupled to noninflaton scalars \cite{Fujita:2018zbr}. In such models, the scale dependence of $G_{L M}(k)$ can be controlled by the shape of the coupling function or the potential. Even in the absence of higher spin fields, nonzero $G_{2M}$ can also be induced by an inflating solid or elastic medium \cite{Bartolo:2013msa, Bartolo:2014xfa} and fossil gravitational waves \cite{Jeong:2012df,Dai:2013kra}. Due to the nonlinearity of gravity, large scale tides beyond the survey region also leave an anisotropic imprint on the observed power spectrum \cite{Akitsu:2016leq,Akitsu:2017syq,Li:2017qgh,Chiang:2018mau,Akitsu:2019avy} and thus yield the form of eq.~\eqref{eq:Pm_GLM}. Note that eq.~\eqref{eq:Pm_GLM} recovers the usual isotropic power spectrum, $P_m({\bf k}) = \bar{P}_m(k)$, by taking $G_{L \geq 1, M} = 0$. 

After substituting eq.~\eqref{eq:P_homo} into eq.~\eqref{eq:Pi_def}, we simplify the $\hat{k}$, $\hat{s}_1$ and $\hat{s}_2$ integrals by use of eqs.~\eqref{eq:math_Ylm} and \eqref{eq:math_int_calXLL}. We finally obtain
\begin{equation}
  \Pi_{\ell \ell_1 \ell_2 \ell'}^{LM X_1 X_2}(k) 
  = \frac{(4\pi)^2 (-1)^{\ell + \ell_2} h_{\ell_1 \ell_2 \ell'} h_{\ell \ell' L}}{(2\ell_1 + 1)(2\ell_2 + 1)\sqrt{2\ell' + 1} \sqrt{2L + 1} } c_{\ell_1}^{X_1}(k) c_{\ell_2}^{X_2} (k)
  \bar{P}_m(k) G_{L M}(k)
 . \label{eq:Pi_GLM}
\end{equation}
Without loss of generality, according to eq.~\eqref{eq:calP_def}, this can be transformed into the reduced coefficient as
\begin{equation}
  {\cal P}_{\ell \ell_1 \ell_2 \ell'}^{LM X_1 X_2}(k)
  = \bar{\cal P}_{\ell \ell_1 \ell_2 \ell'}^{L  X_1 X_2}(k)
  \frac{G_{LM}(k)}{\sqrt{4\pi}} , \label{eq:calP_GLM}
\end{equation}
where
\begin{equation}
 \bar{\cal P}_{\ell \ell_1 \ell_2 \ell'}^{L X_1 X_2}(k) \equiv \frac{(4\pi)^{2} (-1)^{\ell_2} h_{\ell_1 \ell_2 \ell'}^2 h_{\ell \ell' L}^2}{(2\ell_1 + 1)(2\ell_2 + 1)(2\ell' + 1)\sqrt{2L + 1} }
  c_{\ell_1}^{X_1}(k) c_{\ell_2}^{X_2}(k) \bar{P}_m(k) . \label{eq:barcalP}
\end{equation}


\begin{figure}[t]
  \begin{tabular}{cc} 
    \begin{minipage}{0.5\hsize}
      \begin{center}
        \includegraphics[width=1.\textwidth]{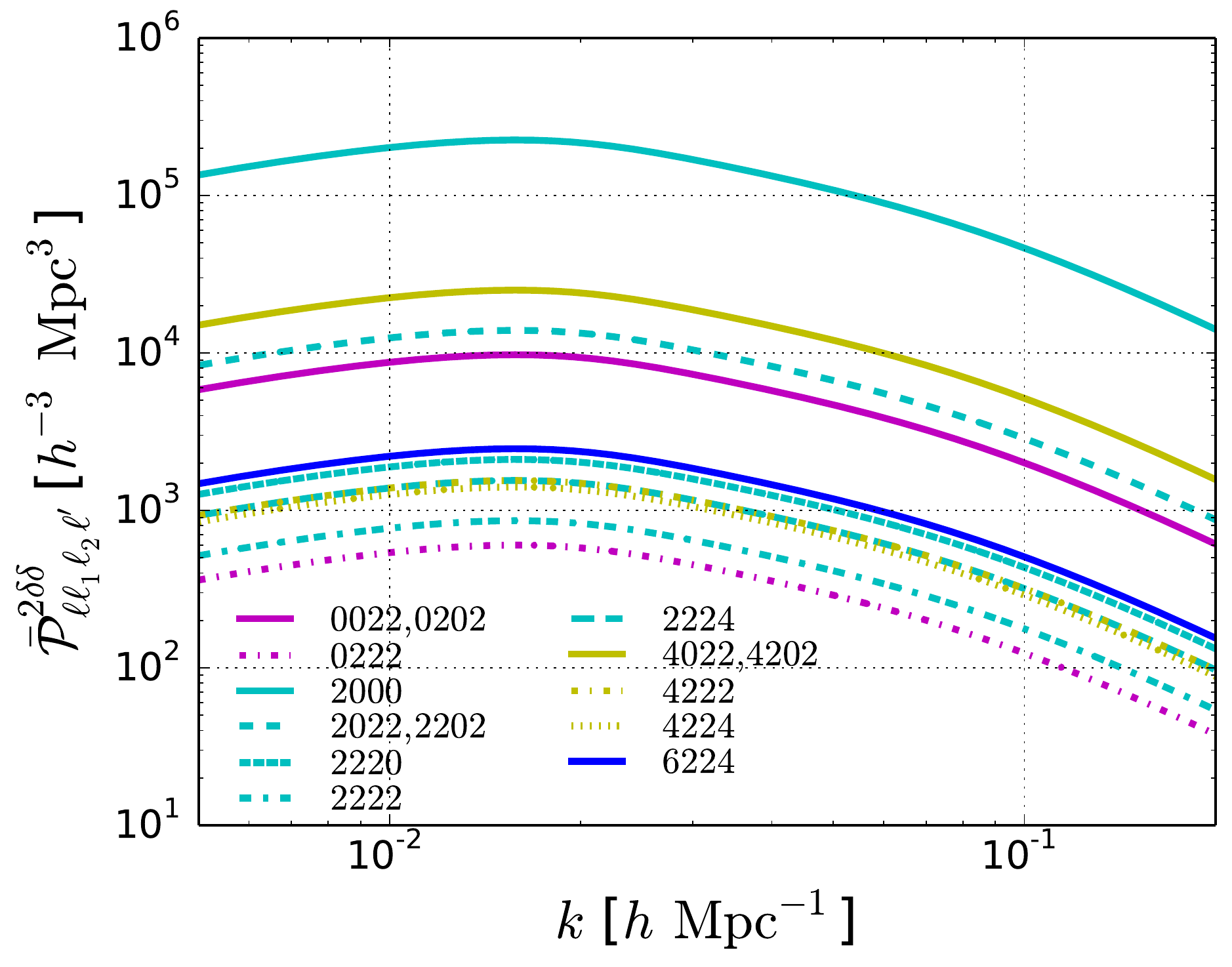}
      \end{center}
    \end{minipage}
    \begin{minipage}{0.5\hsize}
      \begin{center}
        \includegraphics[width=1.\textwidth]{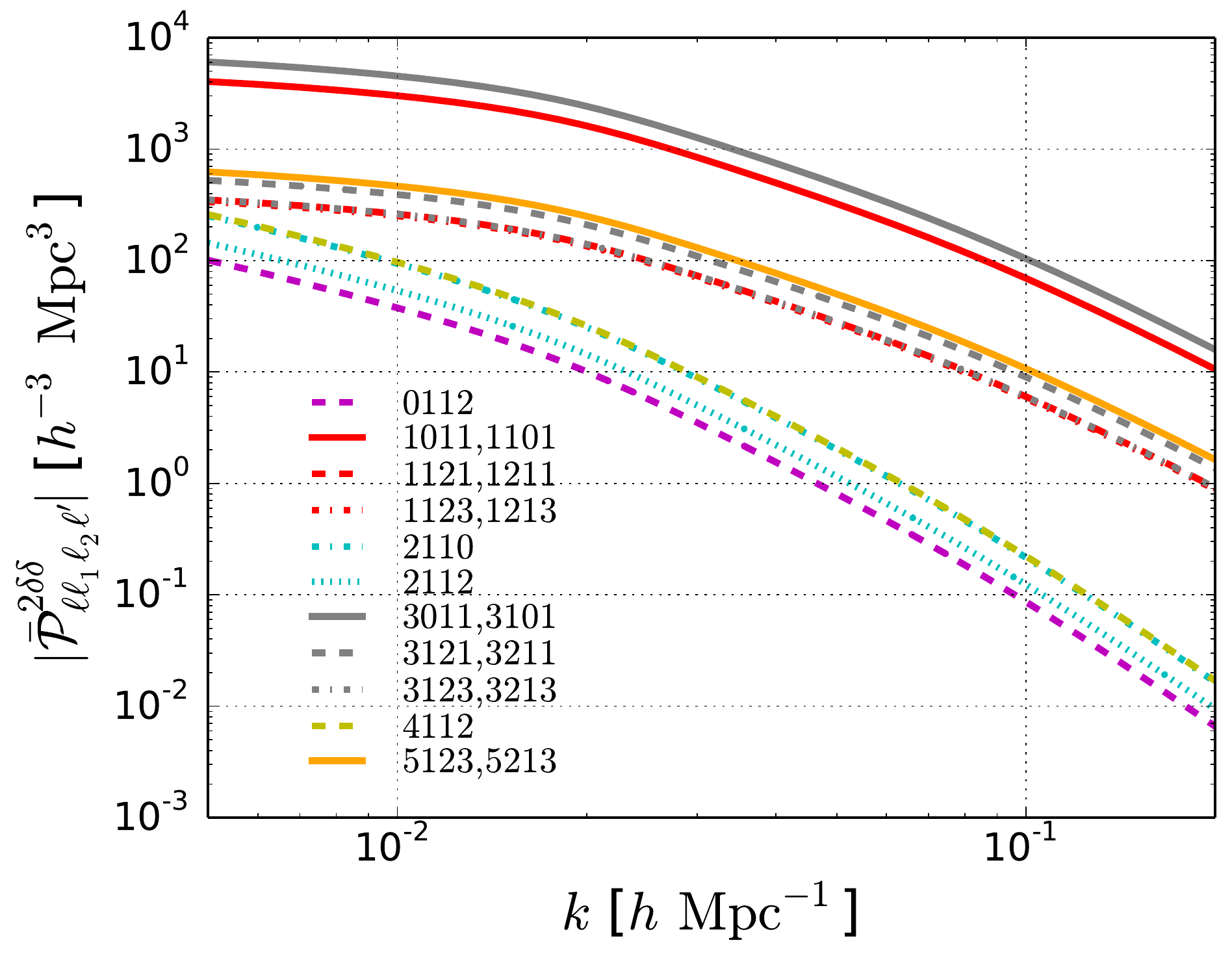}
      \end{center}
    \end{minipage}
  \end{tabular}
  \\
  \begin{tabular}{cc} 
    \begin{minipage}{0.5\hsize}
      \begin{center}
        \includegraphics[width=1.\textwidth]{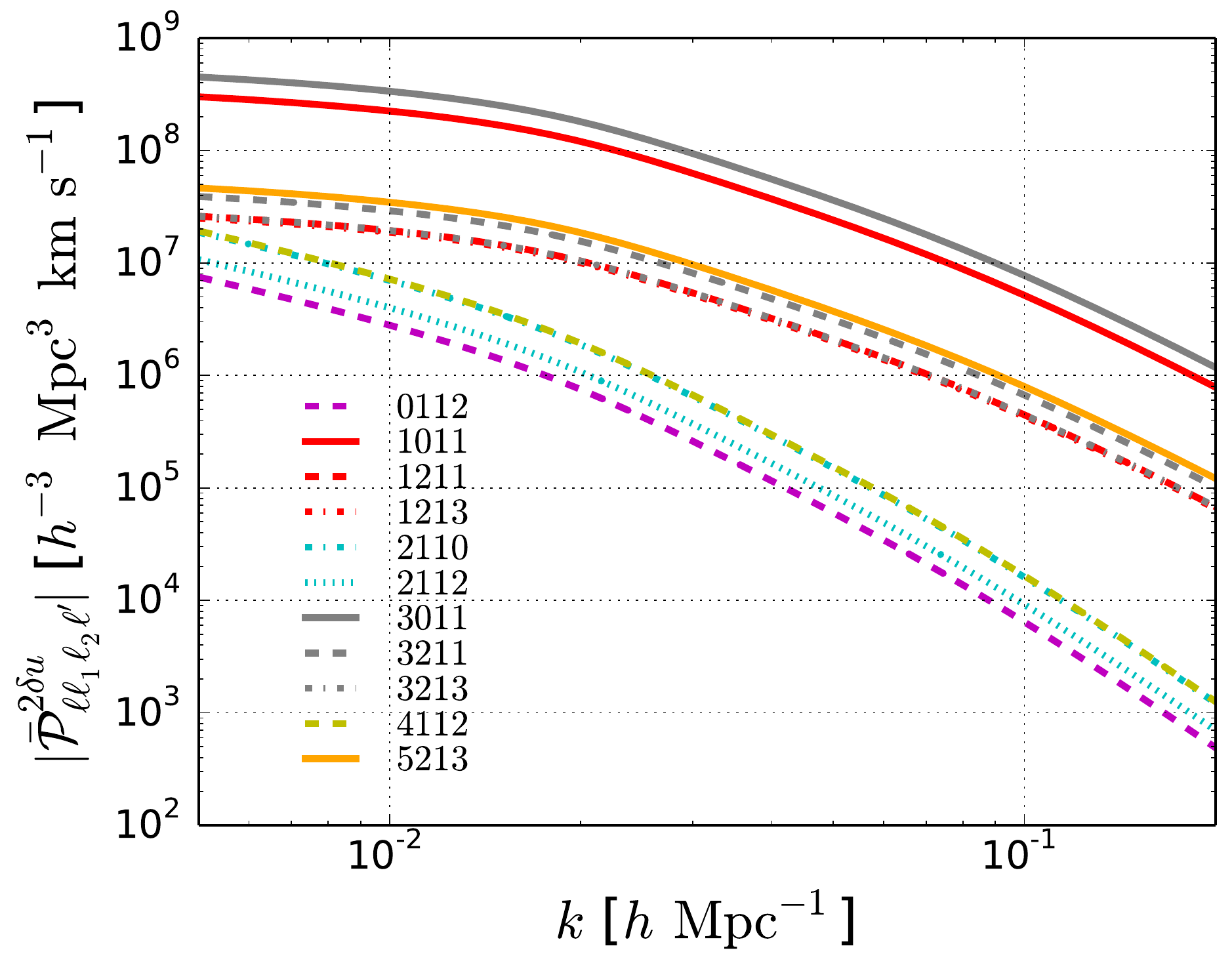}
      \end{center}
    \end{minipage}
    \begin{minipage}{0.5\hsize}
      \begin{center}
        \includegraphics[width=1.\textwidth]{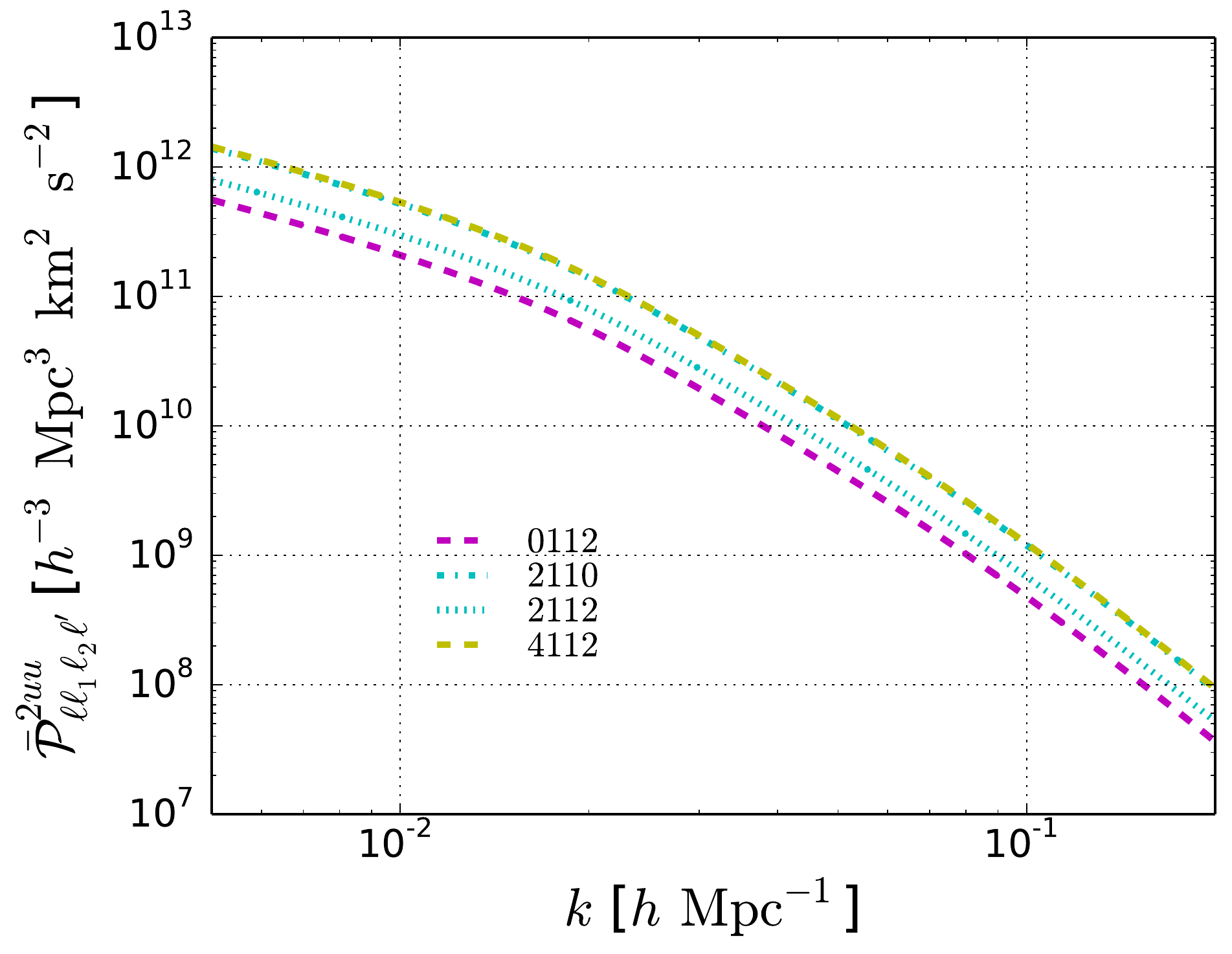}
      \end{center}
    \end{minipage}
  \end{tabular}
  \caption{All nonvanishing components of $\bar{\cal P}_{\ell \ell_1 \ell_2 \ell'}^{2 \delta \delta}$ (top two panels), $\bar{\cal P}_{\ell \ell_1 \ell_2 \ell'}^{2 \delta u}$ (bottom left panel) and $\bar{\cal P}_{\ell \ell_1 \ell_2 \ell'}^{2 uu}$ (bottom right panel), obtained from eq.~\eqref{eq:barcalP}, as a function of $k$ at $b = 2.0$, $z = 0.5$ and $\alpha = 2$. The label of each curve represents the value of $\ell \ell_1 \ell_2 \ell'$. These signals are generated in, e.g., an inflationary model where a spin-1 vector couples to a scalar.}
  \label{fig:calP_L2}
\end{figure}


It is confirmed from this that nonzero $L > 0$ coefficients are the distinctive features of the isotropy-breaking signal. The selection rules of $h_{\ell_1 \ell_2 \ell'}$ and $h_{\ell \ell' L}$ restrict the allowed multipole domain to $|\ell_1 - \ell_2| \leq \ell' \leq \ell_1 + \ell_2$, $|\ell - \ell'| \leq L \leq \ell + \ell'$, $\ell_1 + \ell_2 + \ell' = \rm even$ and $\ell + \ell' + L = \rm even$. The total number of nonvanishing TripoSH coefficients therefore increases with $L$. For example, 32~$\bar{\cal P}_{\ell \ell_1 \ell_2 \ell'}^{2 \delta\delta}$, 11~$\bar{\cal P}_{\ell \ell_1 \ell_2 \ell'}^{2 \delta u}$ and 4~$\bar{\cal P}_{\ell \ell_1 \ell_2 \ell'}^{2 uu}$, and 36~$\bar{\cal P}_{\ell \ell_1 \ell_2 \ell'}^{4 \delta\delta}$, 12~$\bar{\cal P}_{\ell \ell_1 \ell_2 \ell'}^{4 \delta u}$ and 4~$\bar{\cal P}_{\ell \ell_1 \ell_2 \ell'}^{4 uu}$ do not vanish for $L = 2$ and $4$, respectively.

One can see the shapes of $\bar{\cal P}_{\ell \ell_1 \ell_2 \ell'}^{2 \delta \delta}$, $\bar{\cal P}_{\ell \ell_1 \ell_2 \ell'}^{2 \delta u}$ and $\bar{\cal P}_{\ell \ell_1 \ell_2 \ell'}^{2 uu}$ in figure~\ref{fig:calP_L2}. The coefficients sourced by the monopole $c_0^\delta$, corresponding to $\ell_1 = 0$ and/or $\ell_2 = 0$, have the significant amplitudes. The coefficients for $\ell_1 = 1$ and/or $\ell_2 = 1$ have relatively red-tilted shapes because $c_1^{X}(k) \propto k^{-1}$.

\subsection{Covariance}

Here, we compute the covariance of the TripoSH coefficient. The 2PCF reconstructed from a single realization is given by
\begin{equation}
  \hat{\xi}^{X_1 X_2}({\bf s}_{12}, \hat{s}_1, \hat{s}_2) = \int \frac{d^3 k}{(2\pi)^3} e^{i {\bf k} \cdot {\bf s}_{12}} \hat{P}^{X_1 X_2}({\bf k}, \hat{s}_1, \hat{s}_2) ,
  \end{equation}
where
\begin{equation}
  \hat{P}^{X_1 X_2}({\bf k}, \hat{s}_1, \hat{s}_2) \equiv \frac{1}{V} F^{X_1}({\bf k}, \hat{s}_1) F^{X_2}(-{\bf k}, \hat{s}_2) - P_{\rm noise}^{X_1 X_2} .
\end{equation}
with $V$ the survey volume and $P_{\rm noise}^{X_1 X_2}$ the power spectrum of the shot noise. We regard $\xi^{X_1 X_2}$, $P^{X_1 X_2}$ and their decomposition coefficients with hats as quantities estimated from a single realization henceforth.

We assume Gaussianity of $F^X$; thus, the covariance of $\hat{P}^{X_1 X_2}$ can be simplified to
\begin{eqnarray}
 && \Braket{ \hat{P}^{X_1 X_2}({\bf k}, \hat{s}_1, \hat{s}_2) \hat{P}^{\tilde{X}_1 \tilde{X}_2}(\tilde{\bf k}, \hat{\tilde{s}}_1, \hat{\tilde{s}}_2) }_c \nonumber \\ 
  &&\qquad = 4\pi \frac{\delta_{k,\tilde{k}}}{N_k} 
  \left[
     \delta^{(2)}(\hat{k} + \hat{\tilde{k}})
    P_{\rm tot}^{X_1 \tilde{X}_1} ({\bf k}, \hat{s}_1, \hat{\tilde{s}}_1 )
    P_{\rm tot}^{X_2 \tilde{X}_2} (-{\bf k}, \hat{s}_2, \hat{\tilde{s}}_2 ) \right. \nonumber \\
   &&\qquad\qquad\qquad \left.
    + \delta^{(2)}(\hat{k} - \hat{\tilde{k}}) P_{\rm tot}^{X_1 \tilde{X}_2}({\bf k}, \hat{s}_1, \hat{\tilde{s}}_2 )
    P_{\rm tot}^{X_2 \tilde{X}_1}(-{\bf k}, \hat{s}_2, \hat{\tilde{s}}_1 )  \right], \label{eq:P_covmat}
\end{eqnarray}
where $N_k = V k^2 dk / ( 2\pi^2)$ and $P_{\rm tot}^{X_1 X_2} \equiv P^{X_1 X_2} + P_{\rm noise}^{X_1 X_2}$. Besides, supposing that the anisotropic contributions in $P^{X_1 X_2}$ are negligibly small; namely $|G_{L \geq 1, M}| \ll G_{00}$, $P_{\rm tot}^{X_1 X_2}$ is expressed as
\begin{equation}
  P_{\rm tot}^{X_1 X_2} ({\bf k}, \hat{s}_1, \hat{s}_2 )
  = \sum_{j_1 j_2} (-1)^{j_2}
  p_{j_1 j_2}^{X_1 X_2}(k) {\cal L}_{j_1}(\hat{k} \cdot \hat{s}_1) 
   {\cal L}_{j_2}(\hat{k} \cdot \hat{s}_2)  , \label{eq:Ptot}
\end{equation}
where
\begin{equation}
  p_{j_1 j_2}^{X_1 X_2}(k) \equiv c_{j_1}^{X_1}(k) c_{j_2}^{X_2} (k) \bar{P}_m(k) + P_{\rm noise}^{X_1 X_2}  \delta_{j_1, 0} \delta_{j_2, 0} . \label{eq:pj1j2}
\end{equation}

The covariance of $\hat{\Pi}_{\ell \ell_1 \ell_2 \ell'}^{LM X_1 X_2}$ is obtained through the double TripoSH decomposition of the covariance of $\hat{P}^{X_1 X_2}$ as
\begin{eqnarray}
  && \Braket{\hat{\Pi}_{\ell \ell_1 \ell_2 \ell'}^{LM X_1 X_2}(k) \hat{\Pi}_{\tilde{\ell} \tilde{\ell}_1 \tilde{\ell}_2 \tilde{\ell}'}^{\tilde{L} \tilde{M} \tilde{X}_1 \tilde{X}_2}(\tilde{k})  }_c
  = \int d^2 \hat{k}  \int d^2 \hat{s}_1  \int d^2 \hat{s}_2 \,
  \int d^2 \hat{\tilde{k}}  \int d^2 \hat{\tilde{s}}_1  \int d^2 \hat{\tilde{s}}_2 
  \nonumber \\
  && \qquad \times
       {\cal X}_{\ell \ell_1\ell_2 \ell'}^{LM *}(\hat{k},\hat{s}_1,\hat{s}_2)
       {\cal X}_{\tilde{\ell} \tilde{\ell}_1 \tilde{\ell}_2 \tilde{\ell}'}^{\tilde{L} \tilde{M} *}(\hat{\tilde{k}},\hat{\tilde{s}}_1,\hat{\tilde{s}}_2)
      \Braket{ \hat{P}^{X_1 X_2}({\bf k}, \hat{s}_1, \hat{s}_2) \hat{P}^{\tilde{X}_1 \tilde{X}_2}(\tilde{\bf k}, \hat{\tilde{s}}_1, \hat{\tilde{s}}_2) }_c .
\end{eqnarray}
In a similar way to the derivation of eq.~\eqref{eq:Pi_GLM}, we can simplify the $\hat{s}_1$, $\hat{s}_2$, $\hat{\tilde{s}}_1$ and $\hat{\tilde{s}}_2$ integrals by means of eq.~\eqref{eq:math_int_calXLL}. Via the transformation into the reduced coefficient by eq.~\eqref{eq:calP_def}, we derive the form of the covariance matrix of $\hat{\cal P}_{\ell \ell_1 \ell_2 \ell'}^{LM X_1 X_2}$, reading
\begin{eqnarray}
  {\bf C}_{{\cal P} ; {\cal P}^*}
&\equiv&  \Braket{\hat{\cal P}_{\ell \ell_1 \ell_2 \ell'}^{LM X_1 X_2}(k) \hat{\cal P}_{\tilde{\ell} \tilde{\ell}_1 \tilde{\ell}_2 \tilde{\ell}'}^{\tilde{L} \tilde{M} \tilde{X}_1 \tilde{X}_2 *}(\tilde{k})  }_c 
  \nonumber \\ 
  &=& 
  \delta_{L, \tilde{L}} \delta_{M, \tilde{M}} \frac{\delta_{k,\tilde{k}}}{N_k}
  \Upsilon_{\ell \ell_1\ell_2 \ell' ; \tilde{\ell} \tilde{\ell}_1 \tilde{\ell}_2 \tilde{\ell}'}^{L; X_1 X_2 ; \tilde{X}_1 \tilde{X}_2}(k) , \label{eq:calP_covmat}
\end{eqnarray}
where
\begin{eqnarray}
  \Upsilon_{\ell \ell_1\ell_2 \ell' ; \tilde{\ell} \tilde{\ell}_1 \tilde{\ell}_2 \tilde{\ell}'}^{L; X_1 X_2 ; \tilde{X}_1 \tilde{X}_2}(k)
  &\equiv&
  \frac{(4\pi)^4 (-1)^{\ell_2 + \tilde{\ell}_1}  h_{\ell_1 \ell_2 \ell'}^2 h_{\ell \ell' L}^2 h_{\tilde{\ell}_1 \tilde{\ell}_2 \tilde{\ell}'}^2  h_{\tilde{\ell} \tilde{\ell}' L}^2}{(2\ell_1 + 1)(2\ell_2 + 1)(2\ell' + 1)(2 \tilde{\ell}_1 + 1)(2 \tilde{\ell}_2 + 1)(2\tilde{\ell}' + 1)(2 L + 1 )}
  \nonumber \\ 
  && \times \left[ p_{\ell_1 \tilde{\ell}_1}^{X_1 \tilde{X}_1}(k)
    p_{\ell_2 \tilde{\ell}_2}^{X_2 \tilde{X}_2}(k)
    + (-1)^{L}
    p_{\ell_1 \tilde{\ell}_2}^{X_1 \tilde{X}_2}(k)
    p_{\ell_2 \tilde{\ell}_1}^{X_2 \tilde{X}_1}(k)
    \right] , \label{eq:Upsilon}
\end{eqnarray}
with $p_{j_1 j_2}^{X_1 X_2}$ the coefficient of the double Legendre expansion of $P_{\rm tot}^{X_1 X_2}$, defined in eqs.~\eqref{eq:Ptot} and \eqref{eq:pj1j2}. This can be straightforwardly transformed into the covariance of $\hat{\Xi}_{\ell\ell_1\ell_2 \ell'}^{LM X_1 X_2}$ by means of eqs.~\eqref{eq:hankel} and \eqref{eq:calP_def}. 

In analogy with the $L = 0$ case \cite{Shiraishi:2020nnw}, we also find from eqs.~\eqref{eq:Upsilon} and \eqref{eq:pj1j2} that the covariance becomes independent of $P_{\rm noise}^{X_1 X_2} $ at some special multipole configurations, e.g., where none or only one of $\ell_1$, $\ell_2$, $\tilde{\ell}_1$ and $\tilde{\ell}_2$ becomes $0$. The covariance then takes the minimized form as
\begin{equation}
  \Upsilon_{\ell \ell_1\ell_2 \ell' ; \tilde{\ell} \tilde{\ell}_1 \tilde{\ell}_2 \tilde{\ell}'}^{L; X_1 X_2 ; \tilde{X}_1 \tilde{X}_2}(k)
  = \left[ 1 + (-1)^{L} \right]
  \bar{\cal P}_{\ell \ell_1 \ell_2 \ell'}^{L X_1 X_2}(k)
  \bar{\cal P}_{\tilde{\ell} \tilde{\ell}_1 \tilde{\ell}_2 \tilde{\ell}'}^{L \tilde{X}_1 \tilde{X}_2 *}(k).
  \label{eq:Upsilon_CVL}
\end{equation}
This form reveals that there is no entanglement between different multipole moments, which appear in the PP-limit covariance \eqref{eq:Theta} or \eqref{eq:Theta_dd_uu}. This fact results in the improvement of the detectability of $G_{LM}$ as shown in the next section.

\section{Expected errors on $g_{LM}$} \label{sec:error}

In this section, we forecast the detectability of the anisotropic amplitude $G_{L> 0, M}$ for $L = {\rm even}$ by computing the Fisher matrix. We assume that $G_{LM}$ has a power-law shape,
\begin{equation}
  G_{LM}(k) = g_{LM} \left(\frac{k}{k_*} \right)^q,  \label{eq:gLM_def}
\end{equation}
with $k_* \equiv 0.05 \, {\rm Mpc}^{-1}$ the pivot scale adopted in the CMB analysis \cite{Ade:2015hxq,Ade:2015lrj,Akrami:2018odb}. The power-law shape is naturally predicted, e.g, in vector inflation models \cite{Bartolo:2012sd,Bartolo:2014hwa,Bartolo:2015dga}. We then compute $1 \sigma$ errors on $g_{LM}$ for five degrees of tilt: $q = \pm 2$, $\pm 1$ and $0$.

\subsection{Fisher matrix formalism}

The Fisher matrix for $g_{LM}$ is defined as
\begin{equation}
  F_{g_{LM}, g_{\tilde{L}\tilde{M}}} \equiv \frac{\partial \boldsymbol{\cal P}^*}{\partial g_{LM}^*} \,
{\bf C}^{-1}_{{\cal P}^* ; {\cal P} } \,
  \frac{\partial \boldsymbol{\cal P}^\mathsf{T} }{\partial g_{\tilde{L}\tilde{M}}} ,
\end{equation}
where $\boldsymbol{\cal P}$ is a vector composed of the TripoSH coefficients, and $\boldsymbol{\cal P}^\mathsf{T}$ is its transpose. Since the covariance matrix~\eqref{eq:calP_covmat} is diagonalized with respect to $L$ and $M$, the Fisher matrix is also diagonalized as
\begin{equation}
  F_{g_{LM}, g_{\tilde{L}\tilde{M}}} = {\cal F}_L \delta_{L, \tilde{L}} \delta_{M, \tilde{M}}.
\end{equation}
Taking the continuous limit $\sum_k \Delta k \to \int dk$, we obtain
  \begin{equation}
  {\cal F}_{L} = \sum_{\substack{X_1 X_2 \\ \tilde{X}_1 \tilde{X}_2}} 
   V \int_{k_{\rm min}}^{k_{\rm max}} \frac{k^2 dk}{ (2\pi)^3} \left( \frac{k}{k_*} \right)^{2q} 
   \sum_{\substack{\ell \ell_1 \ell_2 \ell' \\ \tilde{\ell} \tilde{\ell}_1 \tilde{\ell}_2 \tilde{\ell}'}}  
\bar{\cal P}_{\ell \ell_1 \ell_2 \ell'}^{L X_1 X_2 *}(k)
   (\Upsilon^{-1})_{\ell \ell_1\ell_2 \ell' ; \tilde{\ell} \tilde{\ell}_1 \tilde{\ell}_2 \tilde{\ell}'}^{L ; X_1 X_2 ; \tilde{X}_1 \tilde{X}_2}(k)
\bar{\cal P}_{\tilde{\ell} \tilde{\ell}_1 \tilde{\ell}_2 \tilde{\ell}'}^{L \tilde{X}_1 \tilde{X}_2}(k), \label{eq:Fish_calP}
  \end{equation}
  where $\sum_{\substack{X_1 X_2 \\ \tilde{X}_1 \tilde{X}_2}}$ and $\sum_{\substack{\ell \ell_1 \ell_2 \ell' \\ \tilde{\ell} \tilde{\ell}_1 \tilde{\ell}_2 \tilde{\ell}'}}$ should be performed in terms of all observable fields and multipoles one wants to take into account, respectively. One can also compute the Fisher matrix based on $\Xi_{\ell \ell_1 \ell_2 \ell'}^{LM X_1 X_2 }$, while a similar result should be obtained since it is related to eq.~\eqref{eq:Fish_calP} via the simple Hankel transformation \eqref{eq:hankel}. The expected $1\sigma$ error is computed according to $\Delta g_{LM} = 1 / \sqrt{{\cal F}_L}$. Numerical results are described in the next subsection.
   
In the simplest case where one uses the information of only one of the three different 2PCFs ($X_1X_2=\{ \delta\delta, \delta u, uu\} $) and a single multipole set, the Fisher matrix is given by
\begin{equation}
  {\cal F}_L|_{\ell \ell_1 \ell_2 \ell'}^{X_1 X_2} =
  V \int_{k_{\rm min}}^{k_{\rm max}} \frac{k^2 dk}{ (2\pi)^3} \left( \frac{k}{k_*} \right)^{2q}   
  \frac{|\bar{\cal P}_{\ell \ell_1 \ell_2 \ell'}^{L X_1 X_2}(k)|^2}{\Upsilon_{\ell \ell_1\ell_2 \ell' ; \ell \ell_1 \ell_2 \ell'}^{L ; X_1 X_2 ; X_1 X_2}(k)} .
\end{equation}
In particular, for $\ell_1 \neq 0$ and $\ell_2 \neq 0$, the covariance takes the minimized form~\eqref{eq:Upsilon_CVL}. The Fisher matrix is accordingly maximized and can be analytically computed as
\begin{equation}
  {\cal F}_L|_{\ell \ell_1 \ell_2 \ell'}^{X_1 X_2}
  = \frac{V k_*^3}{ 16\pi^3} \times
 \begin{cases}
      \left[\left(\frac{k_{\rm max}}{k_*} \right)^{2q+3} - \left(\frac{k_{\rm min}}{k_*} \right)^{2q+3} \right] \frac{1}{2q+3}   &: q \neq -\frac{3}{2} \\
      \ln\left(\frac{k_{\rm max}}{k_{\rm min}}\right)   &: q = -\frac{3}{2}
    \end{cases}
  .  \label{eq:Fish_calP_CVL}
\end{equation}
For example, for $L = 2$, the following $21 + 9 + 4 $ components take this form,
\begin{eqnarray}
  \begin{split}
    & \begin{cases}
        {\cal F}_2|_{0112}^{\delta \delta} , \ \  
        {\cal F}_2|_{0222}^{\delta \delta} , \\
        {\cal F}_2|_{1121}^{\delta \delta} , \ \
        {\cal F}_2|_{1123}^{\delta \delta} , \ \
        {\cal F}_2|_{1211}^{\delta \delta} , \ \
        {\cal F}_2|_{1213}^{\delta \delta} , \\ 
        {\cal F}_2|_{2110}^{\delta \delta} , \ \ 
        {\cal F}_2|_{2112}^{\delta \delta} , \ \ 
        {\cal F}_2|_{2220}^{\delta \delta} , \ \ 
        {\cal F}_2|_{2222}^{\delta \delta} , \ \
        {\cal F}_2|_{2224}^{\delta \delta} , \\
        {\cal F}_2|_{3121}^{\delta \delta} , \ \
        {\cal F}_2|_{3123}^{\delta \delta} , \ \
        {\cal F}_2|_{3211}^{\delta \delta} , \ \
        {\cal F}_2|_{3213}^{\delta \delta} , \\ 
        {\cal F}_2|_{4112}^{\delta \delta} , \ \
        {\cal F}_2|_{4222}^{\delta \delta} , \ \
        {\cal F}_2|_{4224}^{\delta \delta} , \\
        {\cal F}_2|_{5123}^{\delta \delta} , \ \
        {\cal F}_2|_{5213}^{\delta \delta} , \\
        {\cal F}_2|_{6224}^{\delta \delta} ,
      \end{cases}
    \\
    & \begin{cases}
        {\cal F}_2|_{0112}^{\delta u} , \\
        {\cal F}_2|_{1211}^{\delta u} , \ \
        {\cal F}_2|_{1213}^{\delta u} , \\
        {\cal F}_2|_{2110}^{\delta u} , \ \
        {\cal F}_2|_{2112}^{\delta u} , \\
        {\cal F}_2|_{3211}^{\delta u} , \ \
        {\cal F}_2|_{3213}^{\delta u} , \\
        {\cal F}_2|_{4112}^{\delta u} , \\
        {\cal F}_2|_{5213}^{\delta u} , 
      \end{cases}
    \qquad
    \begin{cases} 
      {\cal F}_2|_{0112}^{u u} , \\
      {\cal F}_2|_{2110}^{u u} , \ \
      {\cal F}_2|_{2112}^{u u} , \\
      {\cal F}_2|_{4112}^{u u} .
    \end{cases}
  \end{split}
  \label{eq:eq:Fish_calP_CVL_G2M_list}
\end{eqnarray}
We stress that eq.~\eqref{eq:Fish_calP_CVL} exactly holds irrespective of the shot noise level. In contrast, in the PP-limit analysis \cite{Shiraishi:2016wec,Bartolo:2017sbu}, eq.~\eqref{eq:Fish_calP_CVL} has been approximately derived for $X_1 X_2 = \delta \delta$, though we then must assume a noiseless survey (see the next subsection and appendix~\ref{appen:LPP_error} for details). 

\subsection{Results}

\begin{figure}[t]
\begin{center}
    \includegraphics[width=0.8\textwidth]{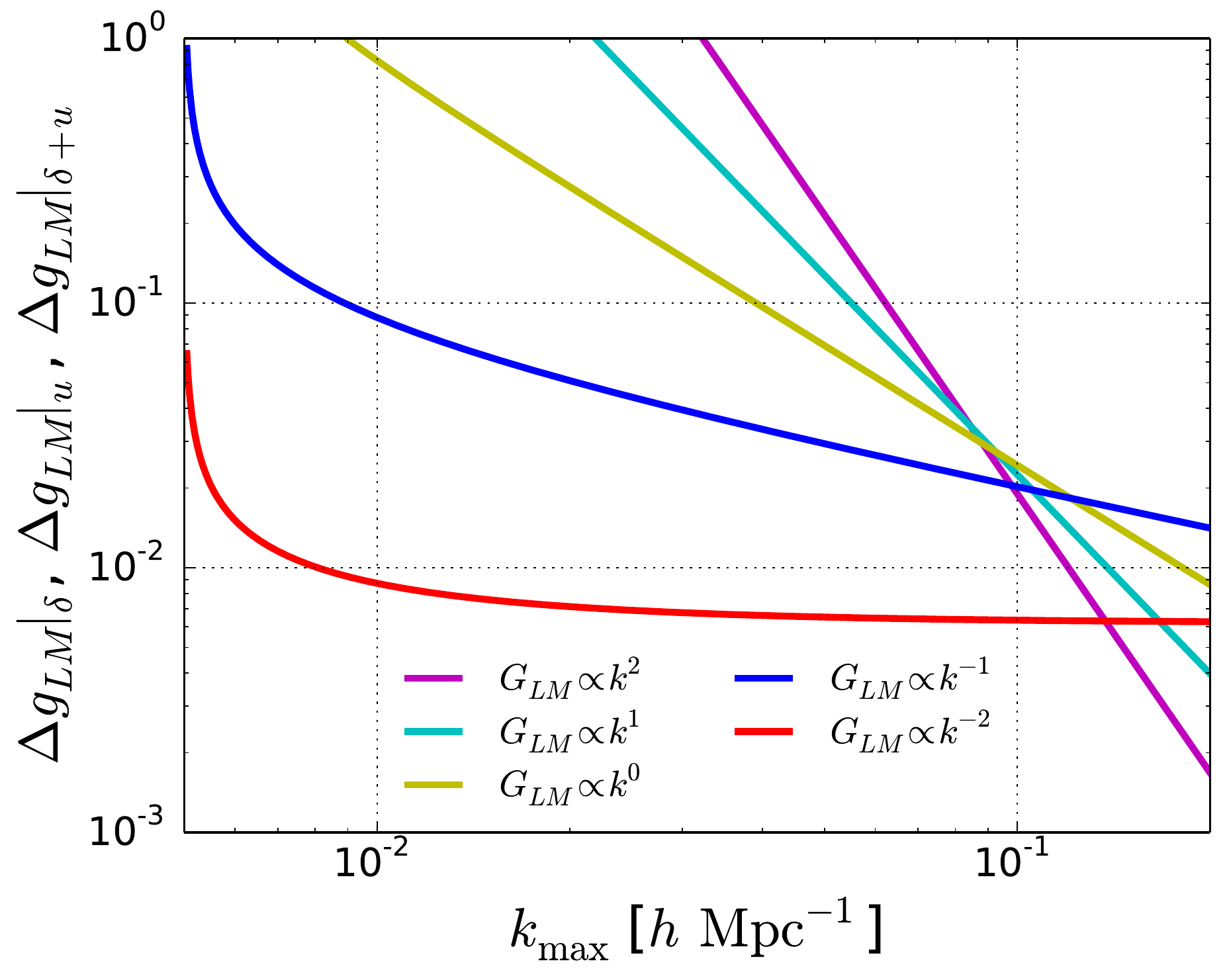}
\end{center}
\caption{Expected $1\sigma$ errors on $g_{LM}$ from the $\delta$-only, $u$-only and $\delta + u$ data as a function of $k_{\rm max}$ when $G_{LM} \propto k^{\pm 2}$, $k^{\pm 1}$ and $k^{ 0}$. These completely overlap with each other. We here adopt $k_{\rm min} = 0.005 h \, \rm Mpc^{-1}$ and $V = 2.5 h^{-3} \, \rm Gpc^3 $.} \label{fig:dgLM}
\end{figure}


\begin{figure}[t]
  \begin{tabular}{cc} 
    \begin{minipage}{0.5\hsize}
      \begin{center}
        \includegraphics[width=1.\textwidth]{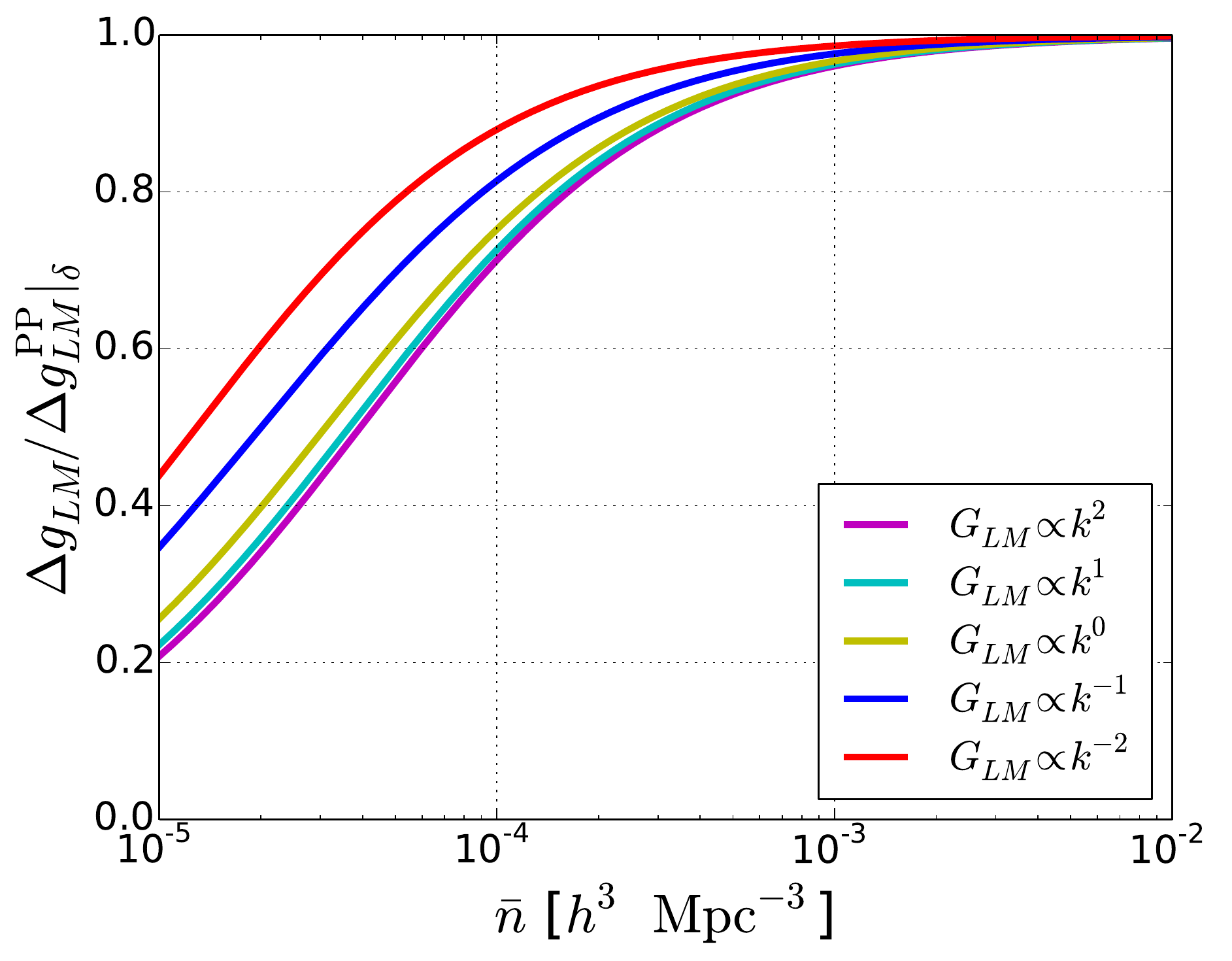}
      \end{center}
    \end{minipage}
    \begin{minipage}{0.5\hsize}
      \begin{center}
        \includegraphics[width=1.\textwidth]{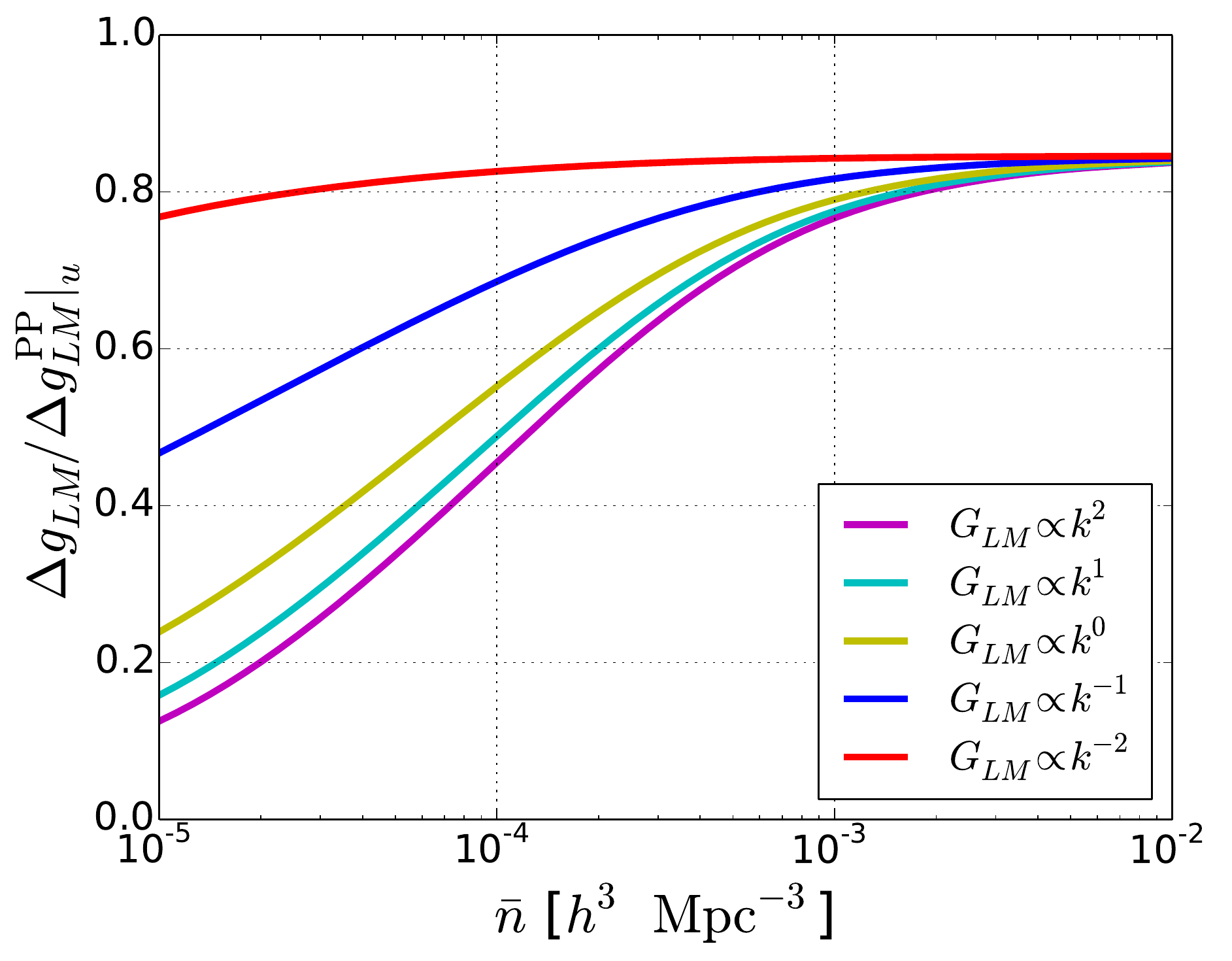}
      \end{center}
    \end{minipage}
  \end{tabular}
  \\
  \begin{tabular}{c} 
    \begin{minipage}{1.\hsize}
      \begin{center}
        \includegraphics[width=0.5\textwidth]{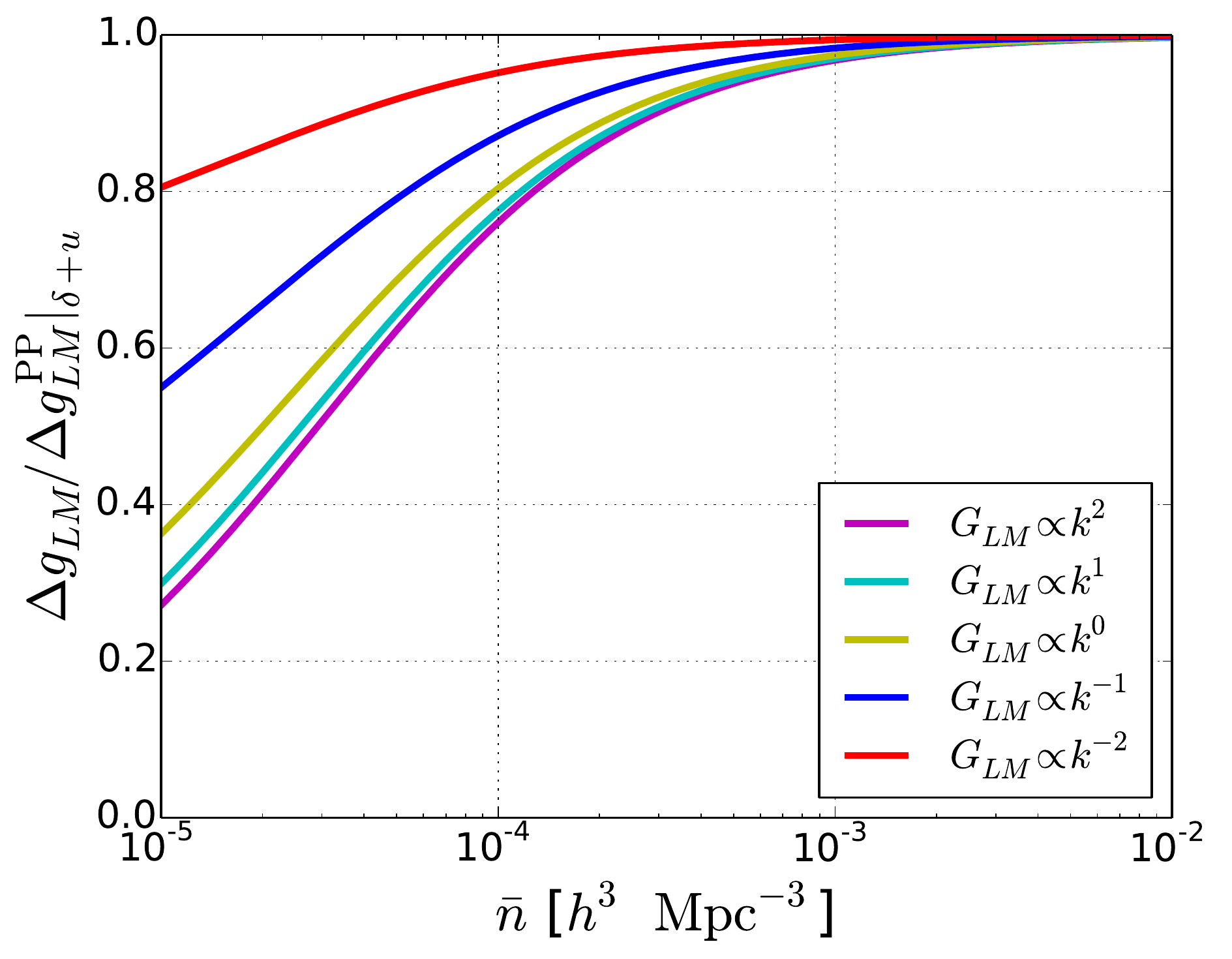}
      \end{center}
    \end{minipage}
  \end{tabular}
  \caption{Ratios of expected $1\sigma$ errors on $g_{LM}$ from the $\delta$-only, $u$-only and $\delta + u$ data to the counterparts estimated in the PP limit as a function of the number density of observed galaxies $\bar{n}$ when $G_{LM} \propto k^{\pm 2}$, $k^{\pm 1}$ and $k^0$. To make this figure, $k_{\rm min} = 0.005 h \, \rm Mpc^{-1}$, $k_{\rm max} = 0.1 h \, \rm Mpc^{-1}$, $b = 2.0$ and $z = 0.5$ are adopted. The results are not so sensitive to the change of these parameters.}
  \label{fig:dgLM_WAbyLPP_per_n}
\end{figure}


Here we discuss numerical results estimated from $\delta$ only ($\Delta g_{LM}|_{\delta}$), $u$ only ($\Delta g_{LM}|_{u}$) and $\delta$ and $u$ jointly ($\Delta g_{LM}|_{\delta + u}$). We then find that $\Delta g_{LM}|_{\delta}$, $\Delta g_{LM}|_{u}$ and $\Delta g_{LM}|_{\delta + u}$ have identical values, and it coincides with the inverse of the root of eq.~\eqref{eq:Fish_calP_CVL}. This means that the corresponding Fisher matrices contain the information of eq.~\eqref{eq:Fish_calP_CVL} in submatrices, while eq.~\eqref{eq:Fish_calP_CVL} already takes the maximized form; thus, there is no additional gain by integrating all multipoles. We also confirm that $\Delta g_{LM}|_{\delta}$, $\Delta g_{LM}|_{u}$ and $\Delta g_{LM}|_{\delta + u}$ are robust against the change of $L$, $M$, $P_{\rm noise}^{X_1 X_2}$, $z$, $b$, $f$ or any cosmological parameter as inferred from eq.~\eqref{eq:Fish_calP_CVL}.

Figure~\ref{fig:dgLM} describes $\Delta g_{LM}|_{\delta}$, $\Delta g_{LM}|_{u}$ and $\Delta g_{LM}|_{\delta + u}$ as a function of $k_{\rm max}$ at the level of current experiments: $k_{\rm min} = 0.005 h \, \rm Mpc^{-1}$ and $V = 2.5 h^{-3} \, \rm Gpc^3$. The $k_{\rm max}$ dependence corresponding to the tilt of $G_{LM}$ is confirmed. The detectability can approach the CMB level $\Delta g_{LM} \sim 10^{-2}$ \cite{Bartolo:2017sbu} as $k_{\rm max}$ is close to $0.1 h \, \rm Mpc^{-1}$. Surpassing this value is expected in a forthcoming survey by increasing $V$. 

Next, we discuss the detectability improvement relative to the PP-limit results $\Delta g_{LM}^{\rm PP}$ computed from eq.~\eqref{eq:Fish_P}. In figure~\ref{fig:dgLM_WAbyLPP_per_n} we show $\Delta g_{LM}/ \Delta g_{LM}^{\rm PP}$ as a function of $\bar{n}$, which determines the sizes of the shot noise spectra according to $P_{\rm noise}^{\delta \delta} = 1 / \bar{n}$, $P_{\rm noise}^{uu} = \sigma_u^2 / \bar{n}$ and $P_{\rm noise}^{\delta u} = P_{\rm noise}^{u \delta} = 0$, where $\sigma_u = 300 \, \rm km/s$ is assumed. Note that this ratio is independent of $V$.

The observed fact that $\Delta g_{LM}/ \Delta g_{LM}^{\rm PP} \leq 1$ for any $\bar{n}$ means that, in detectability, our TripoSH-based analysis always surpasses the PP-limit one. The PP approximation causes nontrivial mode mixings at the covariance level and the degradation of detectability (see appendix~\ref{appen:LPP_covmat} for details). However, the superiority of the TripoSH-based analysis fades in the high number density limit. This is because $\Delta g_{LM}$ remains constant, while $\Delta g_{LM}^{\rm PP}$ shrinks owing to the reduction of the covariance \eqref{eq:Theta}. In the noiseless limit (or equivalently, in the large $\bar{n}$ regime), $\Delta g_{LM}/ \Delta g_{LM}^{\rm PP}|_\delta$ reaches $\sim 1$, while, interestingly, $\Delta g_{LM}/ \Delta g_{LM}^{\rm PP}|_{u}$ still remains $\sim 0.8$, meaning that at least $\sim 20\%$ improvement is always expected compared to the PP-limit analysis regardless of the shape of $G_{LM}$. This benefits from a fact that $\Delta g_{LM}^{\rm PP}|_{u}$ is not minimized even in the noiseless limit due to residual mode mixings in the PP-limit covariance (see appendix~\ref{appen:LPP_error} for a detailed explanation). Comparing eq.~\eqref{eq:Fish_calP_CVL} with the PP-limit analytic results \eqref{eq:Fish_P_d_noiseless} and \eqref{eq:Fish_P_u_noiseless}, these values are recovered.

On the other hand, the superiority of the TripoSH-based analysis becomes remarkable as $G_{LM}(k)$ is blue-tilted. In the blue-tilted case the bias due to the shot noise is more effective at larger $k$ and hence $\Delta g_{LM}^{\rm PP}$ is further smoothed. In contrast, $\Delta g_{LM}$ is free from this effect (because of no dependence on $\bar{n}$) and therefore the gap between them is widened.

\section{Conclusions} \label{sec:conclusion}

In this paper, we, for the first time, have examined the detectability of cosmic statistical anisotropy via galaxy density and velocity surveys without assuming the PP approximation. To extract the anisotropic signal from the 2PCF of the density and velocity fields, we have performed a TripoSH decomposition. The covariance of the TripoSH coefficient has been computed for the first time, and from the resultant analytic expression, the absence of nontrivial mixings between each multipole moment, which appear in the PP-limit covariance, has been confirmed. Owing to this fact, the covariance is minimized, resulting in the optimization of signal detectability. Via a Fisher matrix calculation, we have shown that, in terms of the detectability of statistical anisotropy, our TripoSH-based estimation always has an advantage over the previous PP-limit one. Remarkable superiority has been confirmed, especially in the cases that the shot noise level is large and that target statistical anisotropy has a blue-tilted shape in Fourier space. 

In upcoming all-sky surveys \cite{Dore:2014cca,Laureijs:2011gra,Spergel:2013tha}, wide-angle effects in the 2PCF, which cannot be treated by the PP-limit formalism, will play a critical role. As we have shown in this paper, inclusion of wide-angle effects in the data analysis is beneficial for the cosmic isotropy test. Since the TripoSH decomposition is applicable as long as two LOS directions are distinguishable from each other, an analysis of no all-sky survey data could even achieve a similar detectability improvement.
As some experiments will start running in the few years \cite{Dore:2014cca,Aghamousa:2016zmz}, a feasible estimator of the TripoSH coefficient and a methodology for removing artificial anisotropic contaminations due to, e.g., asymmetric survey geometry should be immediately established.

Throughout this paper, linear theory has always been adopted, and any general relativistic effect (e.g.~\cite{Bertacca:2012tp}) has not been taken into account. Hence, the reasonability of our estimates is not trivial for extremely large or small scales. The geometrical distortion produced by the Alcock-Paczynski effect \cite{Alcock:1979mp,Ballinger:1996cd,Matsubara:1996nf}, has also not been included in our analysis since the geometrical distortion on the curved sky is not trivial. These should be clarified in future works.

\acknowledgments

M.\,S. is supported by JSPS KAKENHI Grant Nos.~JP19K14718 and JP20H05859. T.\,O. acknowledges support from the Ministry of Science and Technology of Taiwan under Grant Nos. MOST 106-2119-M-001-031-MY3 and MOST 109-2112-M-001-027- and the Career Development Award, Academia Sinica (AS-CDA-108-M02) for the period of 2019 to 2023. K.\,A. is supported by Grand-in-Aid for JSPS fellows No.~JP19J12254. M.\,S. and K.\,A. also acknowledge the Center for Computational Astrophysics, National Astronomical Observatory of Japan, for providing the computing resources of the Cray XC50. 



\appendix

\section{Expected errors on $g_{LM}$ in the plane-parallel limit} \label{appen:LPP}

In this appendix, we discuss the Fisher matrix forecasts for $g_{LM}$ in the PP limit ($\hat{s}_1 = \hat{s}_2$). As for the $\delta$-only case, there already exist complete discussions in the literature \cite{Shiraishi:2016wec,Bartolo:2017sbu,Akitsu:2019avy}, while here it is extended by adding the velocity field.

\subsection{Bipolar spherical harmonic decomposition} \label{appen:LPP_BipoSH}

In the PP limit, the 2PCF is characterized by two angles: $\hat{s}_{12}$ and $\hat{s} \equiv \hat{s}_1 = \hat{s}_2$. In this case, the angular dependence is completely decomposed according to
\begin{equation}
 \xi^{X_1 X_2}({\bf s}_{12}, \hat{s}, \hat{s}) = \sum_{\ell \ell' LM} \xi_{\ell \ell'}^{LM X_1 X_2}(s_{12}) X_{\ell \ell'}^{LM}(\hat{s}_{12},\hat{s}) ,
\end{equation}
where we employ the BipoSH basis \cite{Varshalovich:1988ye}, reading
\begin{eqnarray}
  X_{\ell \ell'}^{LM}(\hat{s}_{12},\hat{s})
  &\equiv& \{Y_{\ell}(\hat{s}_{12}) \otimes Y_{\ell'}(\hat{s})\}_{LM} \nonumber \\ 
  &=& \sum_{mm'} {\cal C}_{\ell m \ell' m'}^{LM} Y_{\ell m}(\hat{s}_{12}) Y_{\ell' m'}(\hat{s}) . 
\end{eqnarray}
This is related to the TripoSH basis in the PP limit as
\begin{equation}
  {\cal X}_{\ell \ell_1\ell_2 \ell'}^{LM}(\hat{s}_{12},\hat{s},\hat{s})
  = (-1)^{\ell'} \frac{h_{\ell_1 \ell_2 \ell'}}{\sqrt{2\ell' + 1}} \,
  X_{\ell \ell'}^{LM}(\hat{s}_{12}, \hat{s}).
\end{equation}
The BipoSH coefficient $\xi_{\ell \ell'}^{LM X_1 X_2}$ is related to the Fourier-space counterpart: 
\begin{equation}
\pi_{\ell \ell'}^{LM X_1 X_2}(k) \equiv \int d^2 \hat{k} \int d^2 \hat{s} \, P^{X_1 X_2}({\bf k}, \hat{s}, \hat{s}) X_{\ell \ell'}^{LM *}(\hat{k},\hat{s}) , \label{eq:pi_def} 
\end{equation}
via the Hankel transformation:
\begin{equation}
 \xi_{\ell \ell'}^{LM X_1 X_2}(s_{12}) 
  = i^{\ell} \int_0^\infty \frac{k^2 dk}{2\pi^2}    j_{\ell}(k s_{12})
  \pi_{\ell  \ell'}^{LM X_1 X_2}(k). 
\end{equation}
Note that $\ell$ and $\ell'$ represent the multipole moments associated with $\hat{k}$ (or $\hat{s}_{12}$) and $\hat{s}$, respectively.

Let us define the reduced coefficient as 
\begin{equation}
  P_{\ell \ell'}^{L M X_1 X_2}(k) \equiv (-1)^L \frac{h_{\ell \ell' L}}{\sqrt{4\pi}} 
\pi_{\ell \ell'}^{LM X_1 X_2}(k),
\end{equation}
where $P_{\ell \ell}^{00 X_1 X_2}$ is equivalent to the usual Legendre decomposition coefficient.

\subsection{Anisotropic signal}

Now, we perform the BipoSH decomposition of the redshift-space density and velocity power spectrum originating from the directional-dependent matter power spectrum~\eqref{eq:Pm_GLM}. We rewrite eq.~\eqref{eq:P_homo} into
\begin{equation}
  P^{X_1 X_2}({\bf k}, \hat{s} , \hat{s})
  = \sum_{j} P_{j}^{X_1 X_2}(k)
  {\cal L}_{j}(\hat{k} \cdot \hat{s})
  \sum_{LM} G_{L M}(k) Y_{LM}(\hat{k}), \label{eq:P_homo_LPP} 
\end{equation}
where 
\begin{equation}
  P_{j}^{X_1 X_2}(k) \equiv
\sum_{j_1 j_2} 
\frac{4\pi (-1)^{j_2} h_{j_1 j_2 j}^2 }{(2j_1 + 1)(2j_2 + 1)}
c_{j_1}^{X_1}(k) 
c_{j_2}^{X_2}(k)
\bar{P}_m(k)
\end{equation}
corresponds to the conventional Legendre decomposition coefficient of the isotropic power spectrum. The explicit expressions of all nonzero components read
\begin{eqnarray}
  \begin{split}
P_{0}^{\delta \delta}(k)
&= 
\left(b^2 + \frac{2}{3} bf + \frac{1}{5} f^2 \right)
\bar{P}_m(k), \\
P_{2}^{\delta \delta}(k)
&= \left( \frac{4}{3}bf + \frac{4}{7}  f^2 \right)
\bar{P}_m(k) , \\
P_{4}^{\delta \delta}(k)
&=  \frac{8}{35} f^2 
\bar{P}_m(k), \\
P_{1}^{\delta u}(k) 
&= - P_{1}^{u \delta}(k) 
= - i \left( bf  + \frac{3}{5} f^2 \right) \frac{a H}{k}
\bar{P}_m(k), \\
P_{3}^{\delta u}(k)
&= - P_{3}^{u \delta}(k)
= -\frac{2}{5} i f^2 
\frac{aH}{k} 
\bar{P}_m(k), \\
P_{0}^{u u}(k)
&= \frac{1}{3} f^2 \left( \frac{a H}{k} \right)^2  \bar{P}_m(k), \\
P_{2}^{u u}(k)
&= \frac{2}{3} f^2 \left( \frac{aH}{k} \right)^2   \bar{P}_m(k) ,
\end{split} \label{eq:Leg_coeff_list}
\end{eqnarray}
where we have dropped the terms including $c_1^\delta$ because of the smallness in the PP limit.

Plugging eq.~\eqref{eq:P_homo_LPP} into eq.~\eqref{eq:pi_def} and performing the $\hat{k}$ and $\hat{s}$ integrals lead to
\begin{equation}
  P_{\ell \ell'}^{L M X_1 X_2}(k)
  = 
  \sqrt{\frac{4\pi}{2L+1}}
  \frac{h_{\ell \ell' L}^2}{2\ell' + 1} P_{\ell'}^{X_1 X_2}(k) G_{L M}(k).
\end{equation}
Only the coefficients obeying the selection rule of $h_{\ell \ell' L}$, i.e., $|\ell - \ell'| \leq L \leq \ell + \ell'$ and $\ell + \ell' + L = \rm even$ do not vanish.

\subsection{Covariance} \label{appen:LPP_covmat}

In the PP limit, the covariance of $\hat{P}^{X_1 X_2}$ \eqref{eq:P_covmat} slightly changes as
\begin{eqnarray}
&& \Braket{ \hat{P}^{X_1 X_2}({\bf k}, \hat{s}, \hat{s}) \hat{P}^{\tilde{X}_1 \tilde{X}_2}(\tilde{\bf k}, \hat{\tilde{s}}, \hat{\tilde{s}}) }_c 
  = 4\pi \frac{\delta_{k,\tilde{k}}}{N_k} \times 4\pi \delta^{(2)}(\hat{s} - \hat{\tilde{s}})\nonumber \\ 
&&\qquad\qquad\qquad \times  \left[
     \delta^{(2)}(\hat{k} + \hat{\tilde{k}})
    P_{\rm tot}^{X_1 \tilde{X}_1} ({\bf k}, \hat{s}, \hat{s} )
    P_{\rm tot}^{X_2 \tilde{X}_2} (-{\bf k}, \hat{s}, \hat{s} ) \right. \nonumber \\
   &&\qquad\qquad\qquad\quad \left.
    + \delta^{(2)}(\hat{k} - \hat{\tilde{k}}) P_{\rm tot}^{X_1 \tilde{X}_2}({\bf k}, \hat{s}, \hat{s} )
    P_{\rm tot}^{X_2 \tilde{X}_1}(-{\bf k}, \hat{s}, \hat{s} )  \right]. \label{eq:P_covmat_LPP} 
\end{eqnarray}
In a conventional manner, we expand $P_{\rm tot}^{X_1 X_2}$ using the Legendre polynomials as
\begin{equation} 
  P_{\rm tot}^{X_1 X_2}({\bf k}, \hat{s}, \hat{s} )
  = \sum_{j} P_{{\rm (O)} \, j}^{X_1 X_2}(k)
    {\cal L}_{j}(\hat{k} \cdot \hat{s}) ,
\end{equation}
where
\begin{equation}
  P_{{\rm (O)} \, j}^{X_1 X_2}(k) = 
  P_{j}^{X_1 X_2}(k)
  + P_{\rm noise}^{X_1 X_2} \delta_{j,0} .
  \end{equation}
The double BipoSH decomposition of eq.~\eqref{eq:P_covmat_LPP} leads to
the covariance of $\hat{P}_{\ell \ell'}^{L M X_1 X_2}$ as
\begin{equation}
  \Braket{\hat{P}_{\ell \ell'}^{LM X_1 X_2}(k) \hat{P}_{\tilde{\ell} \tilde{\ell}'}^{\tilde{L}\tilde{M} \tilde{X}_1 \tilde{X}_2 *}(\tilde{k})}_c  
  = \delta_{L,\tilde{L}} \delta_{M,\tilde{M}} \frac{\delta_{k,\tilde{k}}}{N_k} \Theta_{\ell \ell' ; \tilde{\ell} \tilde{\ell}'}^{L; X_1 X_2 ; \tilde{X}_1 \tilde{X}_2}(k) ,
\end{equation}
where
\begin{eqnarray}
  \Theta_{\ell \ell' ; \tilde{\ell} \tilde{\ell}'}^{L; X_1 X_2 ; \tilde{X}_1 \tilde{X}_2}(k)
  &=&  h_{\ell \ell' L}  h_{\tilde{\ell} \tilde{\ell}' L}
  \sum_{j_1 j_2 j} 
    \frac{(4\pi)^3 (-1)^{L + \ell + j_2} h_{j_1 j_2 j}^2 h_{\ell \tilde{\ell} j} h_{\ell' \tilde{\ell}' j} }{(2j_1 + 1)(2j_2 + 1)(2j+1)}
  \left\{
  \begin{matrix}
  L & \tilde{\ell} & \tilde{\ell}' \\
  j & \ell' & \ell
  \end{matrix}
  \right\}
    \nonumber \\ 
&& \times \left[(-1)^{\tilde{\ell}} 
     P_{{\rm (O)} \, j_1}^{X_1 \tilde{X}_1}(k) 
    P_{{\rm (O)} \, j_2}^{X_2 \tilde{X}_2}(k)
    + 
    P_{{\rm (O)} \, j_1}^{X_1 \tilde{X}_2}(k) 
    P_{{\rm (O)} \, j_2}^{X_2 \tilde{X}_1}(k)
    \right]  . \label{eq:Theta}
  \end{eqnarray}
To derive this, we have used eq.~\eqref{eq:math_int_XXLL}. Note that $\Theta_{\ell \ell' ; \tilde{\ell} \tilde{\ell}'}^{L; \delta \delta ; \delta \delta}$ is equivalent to $\Theta_{\ell \ell' , \tilde{\ell} \tilde{\ell}'}^{L}$ in ref.~\cite{Shiraishi:2016wec} although their forms seem different from each other.

Let us focus on the monopole moments $\ell' = \tilde{\ell}' = 0$, which contribute dominantly to the $\delta$-only and $u$-only Fisher matrix. The corresponding elements of the covariance matrix read
\begin{eqnarray}
  \begin{split}
  \Theta_{\ell 0 ; \tilde{\ell} 0}^{L; \delta \delta ; \delta \delta}
  &= \delta_{\ell, L}  \delta_{\tilde{\ell}, L} 
    (2 L+1)
  \left[ 1 + (-1)^{L} \right]
  \left[ (P_{0}^{\delta \delta} + P_{\rm noise}^{\delta \delta})^2 + \frac{1}{5} (P_{2}^{\delta \delta})^2 + \frac{1}{9} (P_{4}^{\delta \delta})^2 \right],  \\
\Theta_{\ell 0 ; \tilde{\ell} 0}^{L; uu ; uu}
    &=  \delta_{\ell, L}  \delta_{\tilde{\ell}, L} (2 L+1) \left[ 1  + (-1)^{L} \right]
\left[ (P_{0}^{uu} + P_{\rm noise}^{uu})^2 + \frac{1}{5} (P_{2}^{uu})^2 \right] .
\end{split} \label{eq:Theta_dd_uu}
  \end{eqnarray}
As confirmed from these, not only the Legendre monopole $P_0^{XX}$ but the other multipoles as $P_2^{XX}$ and $P_4^{XX}$ coexist in the monopole moment of the covariance $\Theta_{\ell 0 ; \tilde{\ell} 0}^{L; XX ; XX}$. The similar mode mixing is also observed for $\ell' > 0$ or $\tilde{\ell}' > 0$. This is due to the identification of four LOS directions by the PP approximation and gives rise to the detectability loss on $g_{LM}$.

\subsection{Fisher matrix forecasts} \label{appen:LPP_error}

The Fisher matrix for $g_{LM}$ in the PP limit reads
\begin{equation}
  {\cal F}_L^{\rm PP} = \sum_{\substack{X_1 X_2 \\ \tilde{X}_1 \tilde{X}_2}} 
V \int_{k_{\rm min}}^{k_{\rm max}} \frac{k^2 dk}{ 2\pi^2}
  \sum_{\substack{\ell \ell' \\ \tilde{\ell} \tilde{\ell}'}}
  \frac{\partial P_{\ell \ell'}^{LM X_1 X_2 *}(k)}{\partial g_{LM}^*}
  (\Theta^{-1})_{\ell \ell' ; \tilde{\ell} \tilde{\ell}'}^{L; X_1 X_2 ; \tilde{X}_1 \tilde{X}_2}(k)
\frac{\partial P_{\tilde{\ell} \tilde{\ell}'}^{LM \tilde{X}_1 \tilde{X}_2}(k) }{\partial g_{LM}} . \label{eq:Fish_P}
\end{equation}
This is used for computation of $\Delta g_{LM} / \Delta g_{LM}^{\rm PP}$  in figure~\ref{fig:dgLM_WAbyLPP_per_n}.

We confirm that ${\cal F}_L^{\rm PP}|_\delta$ and ${\cal F}_L^{\rm PP}|_u$ are mostly determined by the $\ell' = \tilde{\ell}' = 0$ modes and therefore we may evaluate these as
\begin{equation}
  {\cal F}_L^{\rm PP}|_X \simeq V \int_{k_{\rm min}}^{k_{\rm max}} \frac{k^2 dk}{ (2\pi)^3} \left(\frac{k}{k_*} \right)^{2q}
  \frac{(2 L+1) [ P_{0}^{XX}(k)]^2}{\Theta_{L0 ; L0}^{L; XX; XX}(k)} .
\end{equation}

  Let us consider the noiseless limit: $P_{\rm noise}^{X_1 X_2} \ll P_0^{XX}$. As for the $\delta$-only case, the contributions of higher-order moments: $P_{2}^{\delta \delta}$ and $P_{4}^{\delta \delta}$ to the covariance are subdominant; thus, the covariance approximately takes the minimized form:
\begin{equation}
  \Theta_{L0 ; L0}^{L; \delta\delta; \delta\delta}|^{\rm noiseless} \simeq (2L + 1)\left[ 1 + (-1)^{L} \right] (P_0^{\delta\delta})^2.
\end{equation}
Calculating the $k$ integral with this form leads to 
\begin{equation}
  {\cal F}_L^{\rm PP}|_{\delta}^{\rm noiseless} \simeq \frac{V k_*^3}{16\pi^3}  \times
 \begin{cases}
      \left[\left(\frac{k_{\rm max}}{k_*}\right)^{2q+3} - \left(\frac{k_{\rm min}}{k_*}\right)^{2q+3} \right] \frac{1}{2q+3}   &: q \neq -\frac{3}{2} \\
      \ln\left(\frac{k_{\rm max}}{k_{\rm min}}\right)   &: q = -\frac{3}{2}
 \end{cases} , \label{eq:Fish_P_d_noiseless}
\end{equation}
which agrees with the TripoSH-based Fisher matrix~\eqref{eq:Fish_calP_CVL}. On the other hand, as for the $u$-only case, the minimization of the covariance is suppressed as $P_0^{uu}$ and $P_2^{uu}$ contribute comparably to the covariance. Using a fact that $P_2^{uu} = 2P_0^{uu}$ [see eq.~\eqref{eq:Leg_coeff_list}], we derive   
\begin{equation}
  \Theta_{L0 ; L0}^{L; uu; uu}|^{\rm noiseless} = \frac{9}{5} (2L + 1)\left[ 1 + (-1)^{L} \right] (P_0^{uu})^2 ,
\end{equation}
and consequently 
\begin{equation}
  {\cal F}_L^{\rm PP}|_u^{\rm noiseless} \simeq \frac{5}{9} {\cal F}_L^{\rm PP}|_\delta^{\rm noiseless}, \label{eq:Fish_P_u_noiseless}
\end{equation}
meaning that the $\delta$-only case is superior in detectability of $g_{LM}$ to the $u$-only one in the noiseless limit.
These analytic estimates are utilized for explanation of the behavior of $\Delta g_{LM} / \Delta g_{LM}^{\rm PP}$ in the large $\bar{n}$ regime described in figure~\ref{fig:dgLM_WAbyLPP_per_n}.

\section{Useful identities} \label{appen:math}

This section summarizes some mathematical identities used in this paper.

Angular dependences in functions are expanded with the spherical harmonics as
\begin{eqnarray}
  \begin{split}
  {\cal L}_l(\hat{k} \cdot \hat{n}) &= \frac{4\pi}{2l+1} \sum_m  Y_{lm}(\hat{k}) Y_{lm}^*(\hat{n}), 
  \\
  e^{i{\bf k} \cdot {\bf x}} &= \sum_{L M}
  4\pi i^{L}   j_L(kx) Y_{LM}(\hat{k}) Y_{LM}^*(\hat{x}) . 
  \end{split} \label{eq:math_expand}
\end{eqnarray}
Angular integrals of multiples of spherical harmonics can be performed employing 
\begin{eqnarray}
  \begin{split}
  Y_{l_1 m_1}(\hat{n}) Y_{l_2 m_2}(\hat{n}) &= \sum_{l_3 m_3} Y_{l_3 m_3}^*(\hat{n}) h_{l_1 l_2 l_3} \left(
  \begin{matrix}
  l_1 & l_2 & l_3 \\
  m_1 & m_2 & m_3 
  \end{matrix}
 \right), 
 \\
 \int d^2 \hat{n} \, Y_{l_1 m_1}(\hat{n}) Y_{l_2 m_2}^*(\hat{n})
 &= \delta_{l_1, l_2} \delta_{m_1, m_2} . 
 \end{split} \label{eq:math_Ylm}
\end{eqnarray}
Some equations including the Wigner symbols are simplified following
\begin{eqnarray}
  \begin{split}
  \delta_{m_1, m_1'} \delta_{m_2, m_2'} &=
\sum_{l_3 m_3} (2l_3+1)
  \left(
  \begin{matrix}
  l_1 & l_2 & l_3 \\
  m_1 & m_2 & m_3
  \end{matrix}
  \right)
  \left(
  \begin{matrix}
  l_1 & l_2 & l_3 \\
  m_1' & m_2' & m_3 
  \end{matrix}
  \right) , 
  \\
  \frac{\delta_{l_3, l_3'} \delta_{m_3, m_3'}}{2l_3+1} &=
 \sum_{m_1 m_2}
  \left(
  \begin{matrix}
  l_1 & l_2 & l_3 \\
  m_1 & m_2 & m_3
  \end{matrix}
  \right)
  \left(
  \begin{matrix}
  l_1 & l_2 & l_3' \\
  m_1 & m_2 & m_3' 
  \end{matrix}
  \right), 
  \\
  \sqrt{2l + 1}\delta_{L,0} &=
  \sum_{m} (-1)^{l - m}
  \left(   \begin{matrix} l & l & L \\ m & -m & 0\end{matrix}   \right) , 
    \\
    \left(
  \begin{matrix}
  l_1 & l_2 & l_3 \\
  m_1 & m_2 & m_3 
  \end{matrix}
 \right) 
\left\{
  \begin{matrix}
  l_1 & l_2 & l_3 \\
  l_4 & l_5 & l_6 
  \end{matrix}
 \right\} 
 &=   \sum_{m_4 m_5 m_6} (-1)^{\sum_{i=4}^6( l_i - m_i) }
\left(
  \begin{matrix}
  l_5 & l_1 & l_6 \\
  m_5 & -m_1 & -m_6 
  \end{matrix}
  \right) 
  \\
  & \quad \times 
\left(
  \begin{matrix}
  l_6 & l_2 & l_4 \\
  m_6 & -m_2 & -m_4 
  \end{matrix}
  \right)
\left(
  \begin{matrix}
  l_4 & l_3 & l_5 \\
  m_4 & -m_3 & -m_5 
  \end{matrix}
 \right) 
 . 
 \end{split} \label{eq:math_wigner}
 \end{eqnarray}

Employing the above identities, a key angular integral in the computation of the TripoSH coefficient and its covariance can be reduced to
\begin{eqnarray}
  && \int d^2 \hat{s}_1  \int d^2 \hat{s}_2 \,
  {\cal X}_{\ell \ell_1\ell_2 \ell'}^{LM *}(\hat{k},\hat{s}_1,\hat{s}_2)
  {\cal L}_{j_1}(\hat{k} \cdot \hat{s}_1)
  {\cal L}_{j_2}(\hat{k} \cdot \hat{s}_2) \nonumber \\      
  &&\quad= \frac{(4\pi)^2 (-1)^{\ell} h_{\ell_1 \ell_2 \ell'} h_{\ell \ell' L}}{(2\ell_1 + 1)(2\ell_2 + 1)\sqrt{2\ell' + 1} \sqrt{2L + 1} }
   Y_{L M}^*(\hat{k}) 
   \delta_{\ell_1, j_1} \delta_{\ell_2, j_2}
   . \label{eq:math_int_calXLL}
\end{eqnarray}
An angular integral appearing in the computation of the BipoSH covariance is reduced to
\begin{eqnarray}
&&  \int d^2 \hat{k} \int d^2 \hat{s} \,  X_{\ell \ell'}^{LM *}(\hat{k}, \hat{s})  X_{\tilde{\ell} \tilde{\ell}'}^{\tilde{L}\tilde{M} *}(\hat{k}, \hat{s})
  {\cal L}_{j_1}(\hat{k} \cdot \hat{s}) {\cal L}_{j_2}(\hat{k} \cdot \hat{s}) \nonumber \\
  &&\quad =  
  \sum_{j} 
  \frac{(4\pi)^2 (-1)^{\ell + \tilde{\ell}' }  h_{j_1 j_2 j}^2 h_{\ell \tilde{\ell} j} h_{\ell' \tilde{\ell}' j} }{(2j_1 + 1)(2j_2 + 1)(2j+1)}
  \left\{
  \begin{matrix}
  L & \tilde{\ell} & \tilde{\ell}' \\
  j & \ell' & \ell
  \end{matrix}
  \right\}
  (-1)^{M} 
  \delta_{L,\tilde{L}} \delta_{M,-\tilde{M}} . \label{eq:math_int_XXLL}
  \end{eqnarray}


\bibliography{paper}
\end{document}